\documentclass[aps,prb,twocolumn,superscriptaddress,longbibliography,amsmath,amssymb]{revtex4-1}

\usepackage{xcolor}
\usepackage{graphicx}
\usepackage{bm}

\begin{document}

\title{Angular resolved electron energy loss spectroscopy in hexagonal boron nitride}
\author{Fr\'ed\'eric Fossard}
\affiliation{Laboratoire d'Etude des Microstructures, ONERA-CNRS, UMR104, Universit\'e Paris-Saclay, BP 72, 92322 Ch\^atillon Cedex, France}
\author{Lorenzo Sponza}
\affiliation{Laboratoire d'Etude des Microstructures, ONERA-CNRS, UMR104, Universit\'e Paris-Saclay, BP 72, 92322 Ch\^atillon Cedex, France}
\author{L\'eonard Schu\'e}
\affiliation{Laboratoire d'Etude des Microstructures, ONERA-CNRS, UMR104, Universit\'e Paris-Saclay, BP 72, 92322 Ch\^atillon Cedex, France}\affiliation{Groupe d'Etude de la Mati\`ere Condens\'ee, UVSQ-CNRS,  UMR8635, Universit\'e Paris-Saclay, 45 avenue des Etats-Unis, 78035 Versailles Cedex, France}
\author{Claudio Attaccalite}
\affiliation{Aix Marseille University and CNRS, CINAM, UMR7325, Campus de Luminy, case 913, 13288 Marseille, France}
\author{Fran\c{c}ois Ducastelle}
\affiliation{Laboratoire d'Etude des Microstructures, ONERA-CNRS, UMR104, Universit\'e Paris-Saclay, BP 72, 92322 Ch\^atillon Cedex, France}
\author{Julien Barjon}
\affiliation{Groupe d'Etude de la Mati\`ere Condens\'ee, UVSQ-CNRS,  UMR8635, Universit\'e Paris-Saclay, 45 avenue des Etats-Unis, 78035 Versailles Cedex, France}
\author{Annick Loiseau}
\affiliation{Laboratoire d'Etude des Microstructures, ONERA-CNRS, UMR104, Universit\'e Paris-Saclay, BP 72, 92322 Ch\^atillon Cedex, France}

\begin{abstract}
Electron energy loss spectra have been measured on hexagonal boron nitride single crystals employing a novel electron energy loss spectroscopic set-up composed by an electron microscope equipped with a monochromator and an in-column filter. This set-up provides high-quality energy-loss spectra and allows also for the imaging of energy-filtered diffraction patterns. These two acquisition modes provide complementary pieces of information, offering a global view of excitations in reciprocal space. As an example of the capabilities of the method we show how easily the core loss spectra at the $K$ edges of boron and nitrogen can be measured and imaged. Low losses associated to interband and/or plasmon excitations are also measured. This energy range allows us to illustrate that our method provides results of quality comparable to those obtained from non resonant X-ray inelastic scattering, but with advantageous specificities such as an enhanced sensitivity at low $\bm{q}$ and a much higher simplicity and versatility that makes it well adapted to the study of two-dimensional materials and related heterostructures. Finally, by comparing theoretical calculations against our measures, we are able to relate the range of applicability of \textit{ab initio} calculations to the anisotropy of the sample and assess the level of approximation required for a proper simulation of our acquisition method.
\end{abstract}

\date{\today}

\maketitle

\section{Introduction}

Two-dimensional (2D) materials are currently the object of many investigations concerning their electronic and optical properties. Beside and unlike graphene, which is the first representative of this new class of materials, 2D materials are semiconductors with optical properties dominated by excitonic effects which depend on the number of layers and on the nature of the layer stacking.\cite{Xia2014} In this landscape,  $h$-BN  displays a singular situation since it is a large band gap (about 6 eV) semiconductor with a honeycomb lattice similar to that of graphene where boron and nitrogen alternate at the vertices of the honeycomb lattice. Optical measurements on $h$-BN are difficult  because of the necessity  to work in the far UV range and require dedicated laser sources and detection devices. \cite{Watanabe2004,Jaffrennou2007,Watanabe2009,Cassabois2015} Another possibility is to excite the system with electrons and to perform cathodoluminescence experiments.\cite{Jaffrennou2008,Pierret2014,Bourrelier2014, Schue2016} Finally photoemission excitation spectra can be obtained using VUV synchrotron radiation excitation.\cite{Museur2011} All these experiments have clearly shown the importance of excitonic effects in agreement with several theoretical studies \cite{Arnaud2006,Wirtz2006,Wirtz2008,Arnaud2008,Cudazzo2016,Galvani2016} although their exact nature remains far from being fully clarified.  

To go further in understanding excitonic properties, inelastic scattering techniques are  useful and complementary tools to the above cited optical spectroscopies. It is recalled here that the response to electronic excitations is  characterized by the dynamical structure factor $S(\bm{q},\omega)$ which is itself related to the dielectric response $\varepsilon(\bm{q},\omega)$, where $\omega$ and $\bm{q}$ are respectively the energy (or frequency) and momentum variations during the involved scattering process.\cite{Egerton2011,Sturm1993} As far as energy is concerned, optical techniques (absorption, photoluminescence) are very accurate but are confined to the $\bm{q} \to 0$ limit. Recently the full Brillouin zone (and beyond) of $h$-BN single crystals could be explored by means of non resonant inelastic x-ray scattering (NRIXS) experiments,  and energy losses were recorded between a few eV and 40 eV. At low energy the resolution (down to 200 meV) made accessible the investigation of the near edge excitonic regime for different values of $\bm{q}$.\cite{Galambosi2011,Fugallo2015} 

Such experiments can also be performed by using inelastic scattering of fast electrons (electron energy loss spectroscopy (EELS)). This technique has  suffered for a long time of a low energy resolution. This is no longer the case with the latest generations of electron microscopes and these methods can now be used to investigate not only the core loss regime where energy variations are in the range $10^2-10^3$ eV but also the low loss regime, $\omega = 1-50$ eV. The current implantation of electron spectroscopy in transmission electron microscopes makes this technique particularly attractive as it opens the possibility for local investigations at the nanoscale, with no need of large samples, giving access to the impact of defects on the spectroscopic properties. 

In this article we present a novel EELS set-up based on a transmission electron microscope (TEM) tweaked for angular-resolved electron spectroscopy and its application to a detailed study of $h$-BN single crystals. We show that the results are similar to those obtained using synchrotron x-ray sources (NRIXS) in terms of energy resolution, but it exhibits specific advantages: i) It can be employed in two different acquisition modes, also allowing for the measurement of global maps of $S(\bm{q},\omega)$ in the diffraction plane; ii) It has a privileged access to the small-$\bm{q}$ region of the Brillouin zone; iii) The method is fast and can be applied to small samples. This opens the way to a broad field of applications, including 2D materials and their heterostructures. The methods are described in Sec.\ref{experiments}. The results  are presented in Sec. \ref{results}, those for the core loss spectra at the boron  $K$ edge in Sec. \ref{core} and those for the low loss spectra in Sec. \ref{low}. Further documentation can be found in the Supplemental Material.

\section{TEM-EELS experiments}
\label{experiments}

In the past decade, the development of aberration corrected TEM has brought new tools to the scientific community which are particularly suited to image thin materials.\cite{Haider1998,Krivanek1999} Moreover, with the improvement of electron sources and monochromators associated with optimized spectrometers, EELS spectra can be recorded with atomic and sub-eV energy resolutions.\cite{Matijevic2007,Muller2008}  However such systems still make compromises in order to increase the signal by integrating over a finite collecting solid angle\cite{Krivanek1999} or by shining the sample with a very focused beam with a large illumination angle so that the angular dependent information is averaged or truncated.\cite{Suenaga2010} The techniques presented here avoid these disadvantages. They combine energy filtered TEM (EFTEM) and EELS. The electron microscope is a Zeiss Libra 200 MC equiped with an electrostatic CEOS monochromator, an in-column $\Omega$ filter and a Gatan ultrascan 1000 CCD camera. The microscope operates at 80 kV and the monochromated beam gives a resolution of 100 meV with the narrowest slit. The K\"ohler illumination ensures that the beam is parallel, and that its convergence is kept below 80 $\mu$rad.

\begin{figure}[!ht]
\begin{center}
\includegraphics*[width=9cm]{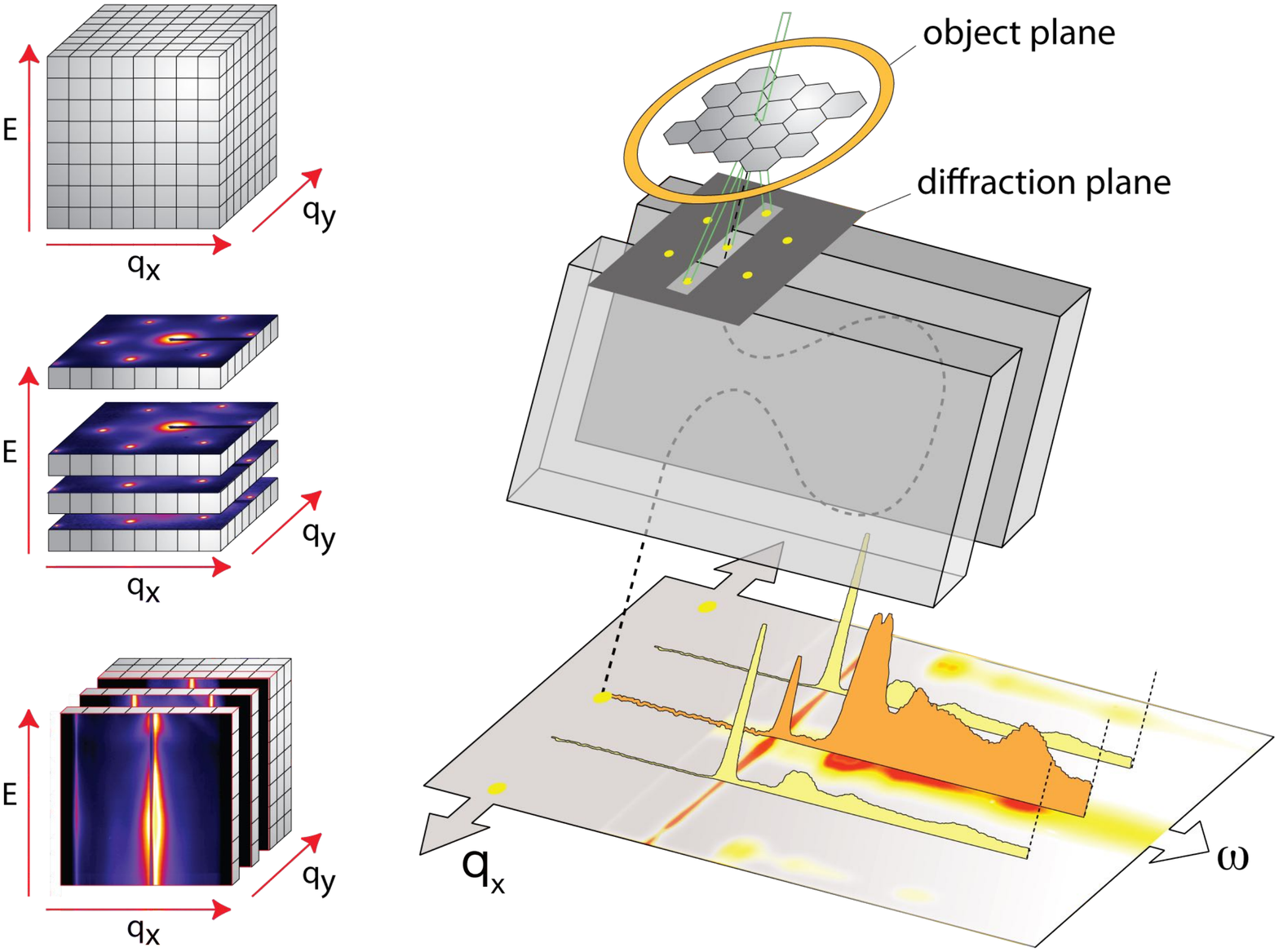} 
\caption{Upper left : datacube in reciprocal space ; middle left : $(E, q_x, q_y)$ datacube construction in EFTEM ; bottom left : $(E, q_x, q_y)$ datacube contruction in $\omega-\bm{q}$ map. Right : schematic principle of the $\omega-\bm{q}$ map acquisition.
}
\label{datacube}
\end{center}
\end{figure}

To measure the dynamical structure factor it is convenient to work within the diffraction plane of the microscope where the scattering angles can be related to the transferred momenta. Since the tranferred momentum is much smaller than the momentum of the incoming beam, the relation between them is given by  $q^2 \simeq k^2(\theta^2 + \theta^2_E)$, where $\bm{k}$ is the initial momentum, $\theta$ is the scattering angle and $\theta_E$ is proportional to the energy loss.\cite{Egerton2011} For a given orientation of the sample, a data cube is built from the values of $\omega, q_x$ and $q_y$, the incident beam being along the $z$ direction.\cite{Jeanguillaume1989}. The component along this direction, $q_E = k\theta_E$ is negligible in general, except when $q$ is close to zero. This is illustrated in Fig.~\ref{datacube}. 

Two strategies have been applied to record this information and are described below.
A first method consists in recording  scattering patterns at given energy losses and to stack them in order to build the horizontal slices of the datacube. The main advantage of this procedure is to obtain the $q_x$ and $q_y$ values of the transferred momentum with the same resolution. The spectral resolution of the EFTEM experiment is determined by the exit slit of the energy filter which selects a bandwidth in the energy-selecting plane.\cite{Reimer1988}  The filtered electrons within this bandwidth form the scattering pattern. The intensity of the signal is an order of magnitude lower than the intensity of a usual diffraction pattern. As a consequence, the integrating time to record one slice is usually larger than 10~s. Data are measured every 0.25 eV  to get a smooth spectrum which is used to subtract a power law background  in the $\omega$ direction of the data cube for every $(q_x, q_y)$ pixel.

In order to take full advantage of the monochromator and to obtain more quantitative results on the double $\omega-\bm{q}$ dependence, a second method is to combine the imaging capabilities of the projective system with the energy dispersion generated by the $\Omega$ energy filter in the microscope. A rectangular slit can be placed at the entrance of the filter in order to select a direction in reciprocal space.\cite{Reimer1989,Wachsmuth2013}  The orientation of the slit is fixed and its larger dimension is perpendicular to the direction of the energy dispersion.  The orientation of the sample must  be adjusted in order to align the slit with a specific crystallographic direction. This can be done using a tilt/rotation holder ensuring a  $360^\circ $ rotation around the optical axis.  Furthermore, since the slit is placed after the first projective system of the microscope, we can also rotate the diffraction pattern by changing the camera length.

The multipoles of the filter must be adjusted in order to keep the $q_x$ information in the direction of the slit while dispersing the energy of the electrons. The image recorded on the camera is therefore in the $(q_x, \omega)$ plane as illustrated in Fig.\ \ref{datacube} (right). Thus, by shifting the scattering pattern with the first projector system, we can scan the $q_y$ direction using discrete steps. In other words, the data cube is now built with vertical slices as illustrated in Fig.\ \ref{datacube} (bottom left).

It is worth noting that the $\bm{q}$-resolution in both experiments  depends on several instrumental parameters as well as on sample specifications. Indeed, both experiments are diffraction experiments and the broadening of the signal is related to the illumination angle, the camera length used, the optical design of the column, the width of slits and apertures.\cite{Wachsmuth2013}One should also take into account the crystallinity of the sample over the diffracting area. In our case, the area is delimitated by an aperture which gives a virtual circular area with a diameter of 70 nm. It will be shown below, in Sec.\ \ref{results}, that the resolution along $q_x$ can be estimated about a few $10^{-2}$ \AA$^{-1}$, whereas the thickness of the slice along $q_y$ in the $\omega-\bm{q}$ mode is about $0.2$ \AA$^{-1}$.

The procedures described above requires $h$-BN samples with well defined orientations. Three slabs have been cut by focused ion beam from a $h$-BN single crystal\cite{Taniguchi2007}  along  $(0001), (10\bar{1}0)$ and $(11\bar{2}0)$ crystallographic planes whose normal directions in the first Brillouin zone  are the $\Gamma A, \Gamma M$ and $\Gamma K$ (see Fig.~\ref{coreloss}). Further details are given in the Supplemental Material.

\section{Results}
\label{results}
\subsection{Core losses at the boron $K$ edge}
\label{core}

Studying electron energy loss (EEL) at the boron K-edge is a textbook case for illustrating the potentialities of our technique. The different peaks are sharp and related to well-known transitions with typical symmetries between the deep $1s$ level and the first unoccupied  $\pi^*$ states (192 eV) and $\sigma^*$ states (199 eV).\cite{Leapman1983,Saito1986,Watanabe1996,Franke1997,Jaouen2000,Arenal2007,Feng2008,McDougall2014} 
\begin{figure}[!ht]
\begin{center}
\includegraphics*[width=8cm]{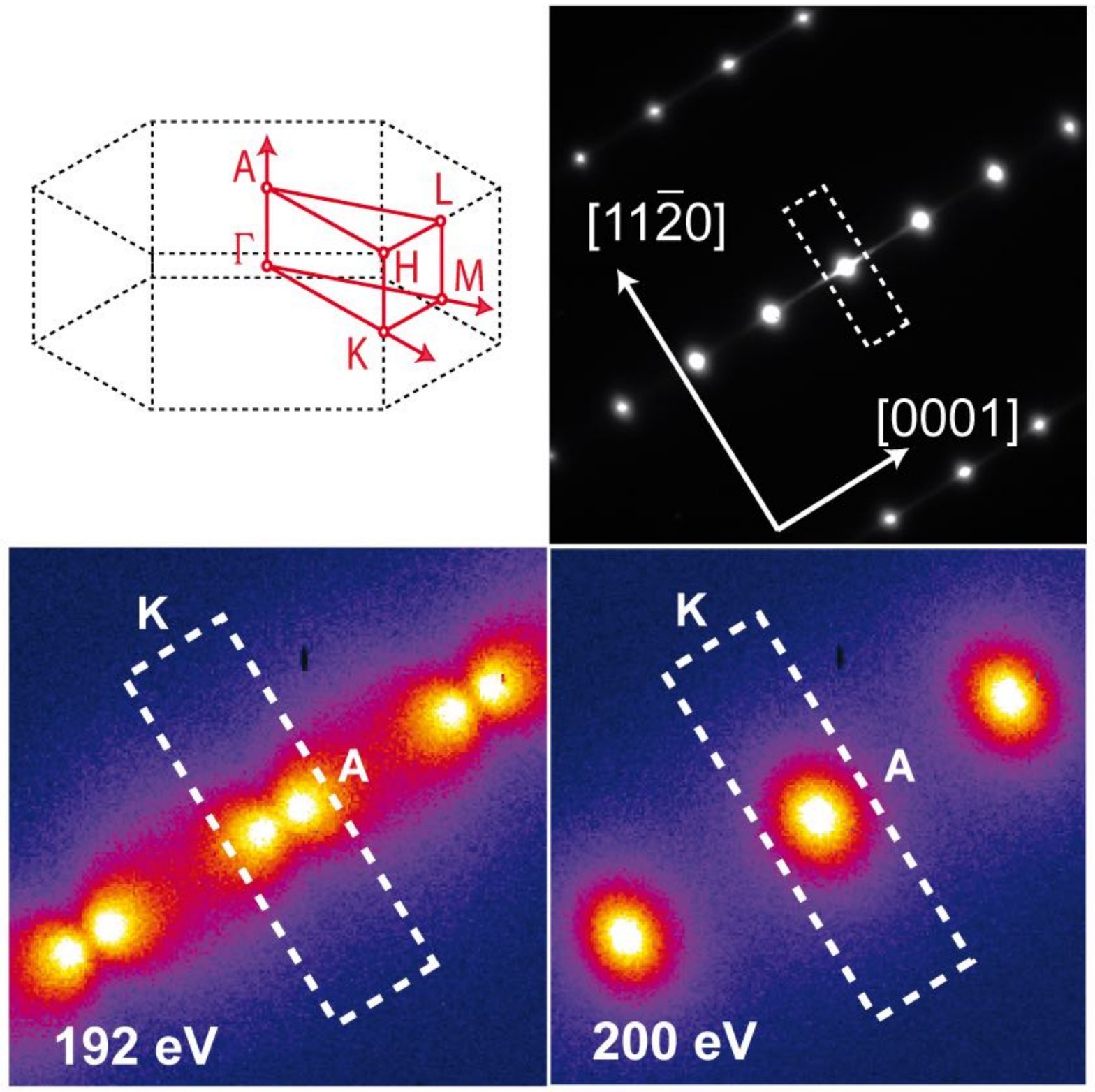} 
\caption{Top: Hexagonal Brillouin zone and diffraction pattern of $h$-BN in a plane containing the hexagonal axis (direction $\Gamma A$) and the $\Gamma K$ direction. Bottom: Energy filtered scattering patterns recorded at 192 eV (left) and at 200 eV (right).
}
\label{coreloss}
\end{center}
\end{figure}
\begin{figure}[!ht]
\begin{center}
\includegraphics*[width=8cm]{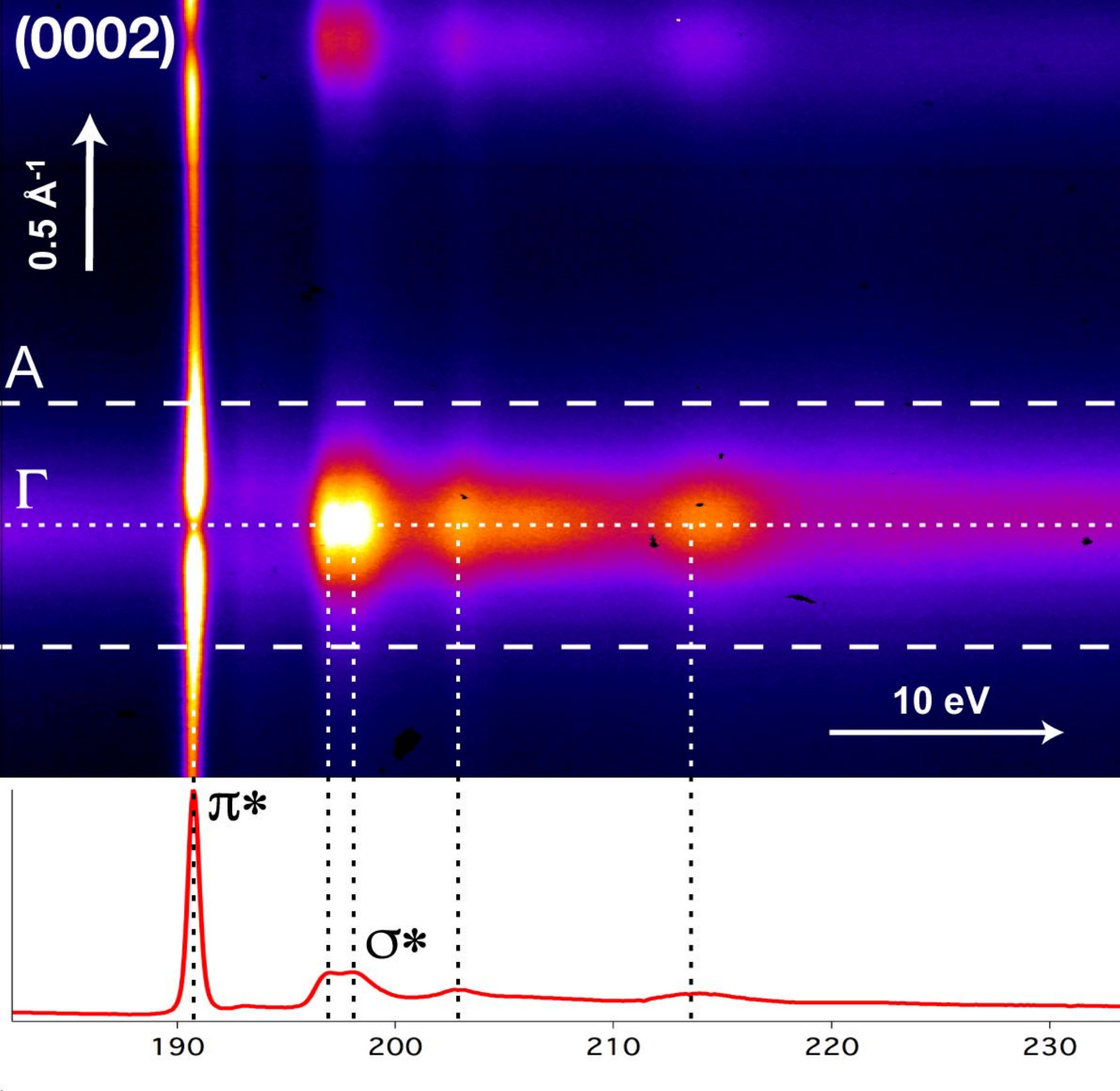} 
\caption{Top: $\omega-\bm{q}$ image recorded in the $\Gamma A$ direction close to the boron $K$-edge energy. Dashed lines delimitate the Brillouin zone. Bottom : related EELS spectrum integrated over the whole Brillouin zone. Dotted lines indicate the Brillouin zone section and significant edge structures. }
\label{core_omega_q}
\end{center}
\end{figure}
Energy filtered scattering patterns have been recorded in the 185-215 eV range for the three samples. Fig.\ \ref{coreloss} presents the elastic diffraction pattern of the second sample with labeled directions as well as  inelastic filtered patterns at the $1s \to \pi^*$ energy (192 eV) and the one close to the $1s \to \sigma^*$ energy (200 eV) obtained by EFTEM.  Notice that in both cases diffuse intensities also occur around the Bragg peaks. This is due to  double-scattering processes involving inelastic scattering and elastic Bragg scattering.  At 192 eV, all  diffraction spots are splitted into two symmetric lobes with the specific [0001] orientation (along the axis of the hexagonal cell). The corresponding $\omega-\bm{q}$ plot in the $\Gamma A$ direction is shown in Fig.\ \ref{core_omega_q}. We have a clear illustration here of the anisotropy of the losses in the $1s \to \pi^*$  transition, which can simply be explained as follows. \cite{Leapman1983,Feng2008} 

\begin{figure}[!ht]
\begin{center}
\includegraphics*[width=7cm]{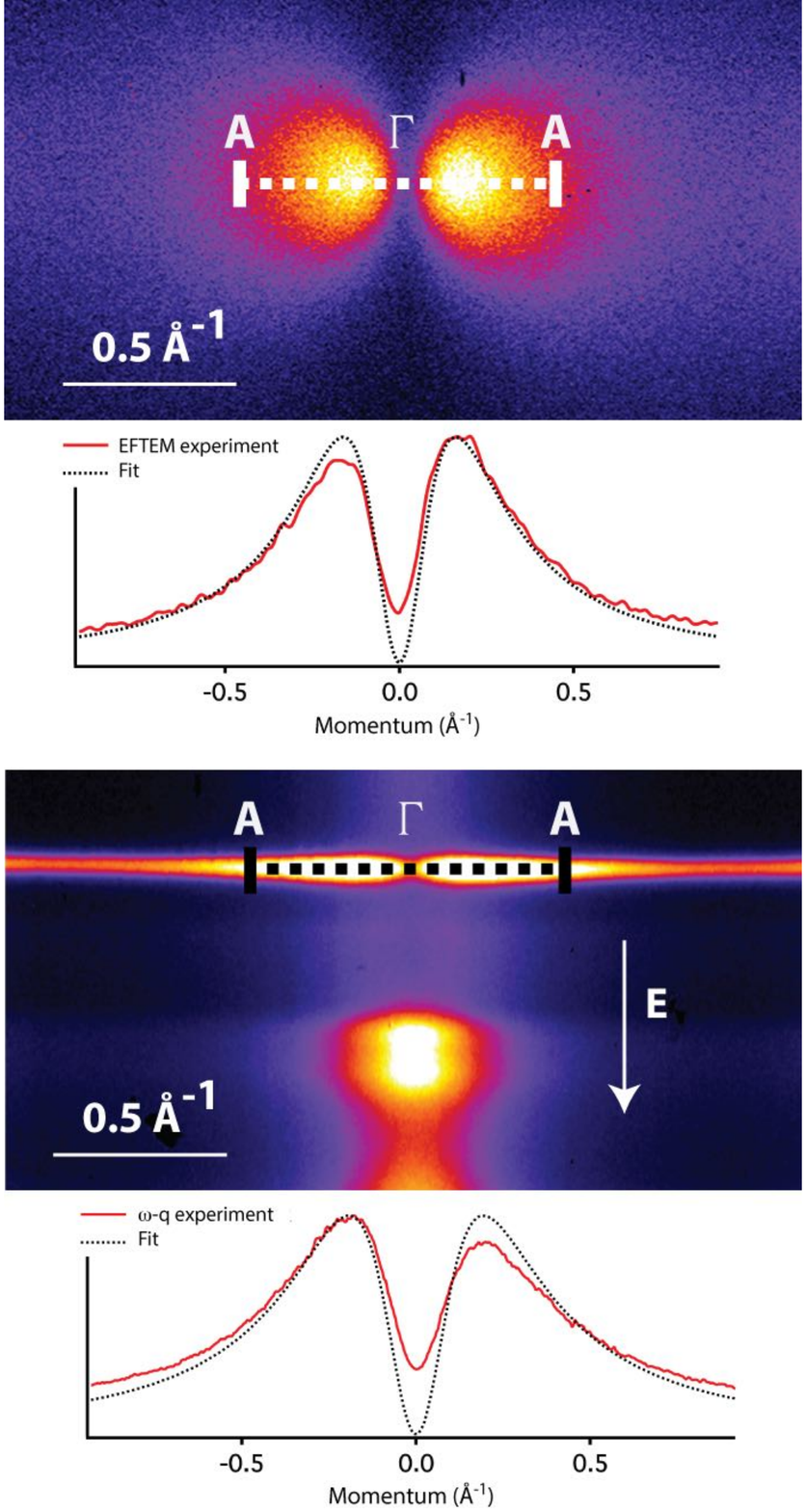} 
\caption{Enlargement of the inelastic scattered intensity corresponding to the $\sigma \to \pi^*$ transition and related profile in the $\Gamma A$ direction in EFTEM at 192 eV (left) and $\omega-\bm{q}$ plot (right). The fit is made using a profile function proportional to $\bar{\bm{q}}^2  / (\bar{\bm{q}}^2 + q_E^2)^2$ with $q_E=0.20$ \AA$^{-1}$.}
\label{qresolution}
\end{center}
\end{figure}

In the simplest single-electron picture the dynamic structure factor is given by:
\begin{equation}
S(\bm{q},\omega) = \sum_f |\langle f|e^{i\bm{q}.\bm{r}} |i \rangle|^2 \delta(E_f-E_i -\hbar\omega) \; ,
\label{dynamic}
\end{equation}
where $i$ and $f$ denote the one-electron initial and final states, respectively, the differential cross section $d^2\sigma/d\Omega\, dE$ for electron scattering being equal to $4S(\bm{q},\omega)/(a_0^2 q^4)$, where $a_0$ is the Bohr radius. The matrix element $\langle f|e^{i\bm{q}.\bm{r}} |i \rangle$ reduces here to the matrix element between the core boron $1s$ function and the conduction band states. In the case of $h$-BN the conduction states at low energy are concentrated on the boron atoms, and more precisely on their $\pi_\parallel$ states pointing along the hexagonal axis,\cite{Galvani2016} so that finally, within the dipolar approximation,  we have to calculate the dipolar matrix element  $\langle \pi_\parallel|\bm{q}.\bm{r} |1s\rangle$. Because of the symmetry of the $\pi_\parallel$ state, only the component of $\bm{r}$ along the hexagonal axis survives, so that we expect that $S(\bm{q},\omega)/q^4 \simeq q_\parallel^2/q^4 =\cos^2\alpha/q^2$, where here $\alpha$ is the angle betwen $\bm{q}$ and the hexagonal axis.\cite{Leapman1983} This means that the symmetry of the scattered intensity around the origin should  be similar to that of the $\pi$ electron density itself. This is clearly the case as shown in Fig.\ \ref{coreloss} and in Fig.\ \ref{qresolution} where an enlargement of the central lobes is shown.

More precisely let us decompose the scattering wave vector $\bm{q}$ into its component in the diffraction plane $\bar{\bm{q}}$ and its inelastic component $\bm{q}_E$ along the incident beam, normal to this plane. \cite{Wachsmuth2013} Then the scattering cross section measured in the diffraction plane is proportional to  $\bar{\bm{q}}^2 \cos^2\alpha / (\bar{\bm{q}}^2 + q_E^2)^2$. This induces an intensity dip along the hexagonal lattice, of width equal  to $2 q_E$, when approaching the origin. Actually, in this limit $\bar{\bm{q}} \to 0$, the scattering vector is normal to the diffraction plane and therefore in the nodal plane of the $\pi$ orbital. We can calculate $q_E$, equal to $E/\hbar v_0$ where $E$ is the energy loss and $v_0$ the electron velocity determined by the accelerating tension of the microscope. In our case, $q_E \simeq 0.20$ \AA$^{-1}$, and it can be seen in Fig.\ \ref{qresolution} that the above formula fits perfectly the measured profiles. This indicates that the (angular) resolution in $q$-space is very good. It is estimated to a few  $10^{-2}$ \AA$^{-1}$.

 At higher energy (200 eV) the diffraction pattern is  modified with an intensity much more isotropic and with an extension in the basal plane typical of the appearance of $\sigma$ states, as also discussed by Leapman et al.\cite{Leapman1983}

The difference between the two regimes is even more obvious when looking at the $\omega-\bm{q}$ plot shown in Fig. \ref{core_omega_q}. It can be noticed that the $\pi^*$ peak at low energy is  separed from a quasi-continuum starting at higher energy in correspondence with the $\sigma^*$ peak, which is typical of an excitonic behaviour. The splitting of the main $\sigma^*$ peak, apparent in the $\omega-\bm{q}$ plot has probably also an excitonic origin.\cite{Wang2015} The corresponding EEL spectrum for $\bar{\bm{q}}$ along the hexagonal axis is shown in Fig.\ \ref{qresolution}. Actually the presence of a core hole is  important here and the single electron description should be improved. This has been done, in particular in Ref. [\onlinecite{Feng2008}] where the authors calculate the full dielectric constant using the Bethe Salpeter formalism\cite{Strinati1988} and found good agreement with the experimental data.

$\omega-\bm{q}$ plots have also been obtained for energies close to the nitrogen K edge (see Supplemental Material). Then the core hole is on the nitrogen atom, but the electron in the conduction band is still concentrated on the boron atoms. Excitonic effects and oscillator strengths are therefore expected to be  weaker, which is the case: The measured  EELS signals are much weaker and then less accurate, and the spectra show broader and less ``atomic-like'' features, in agreement with previous studies.\cite{Franke1997,Watanabe1996,Arenal2007,Pacile2010,Wang2015}

\subsection{Low loss region}
\label{low}

\begin{figure*}[!ht]
\begin{center}
\includegraphics*[width=15cm]{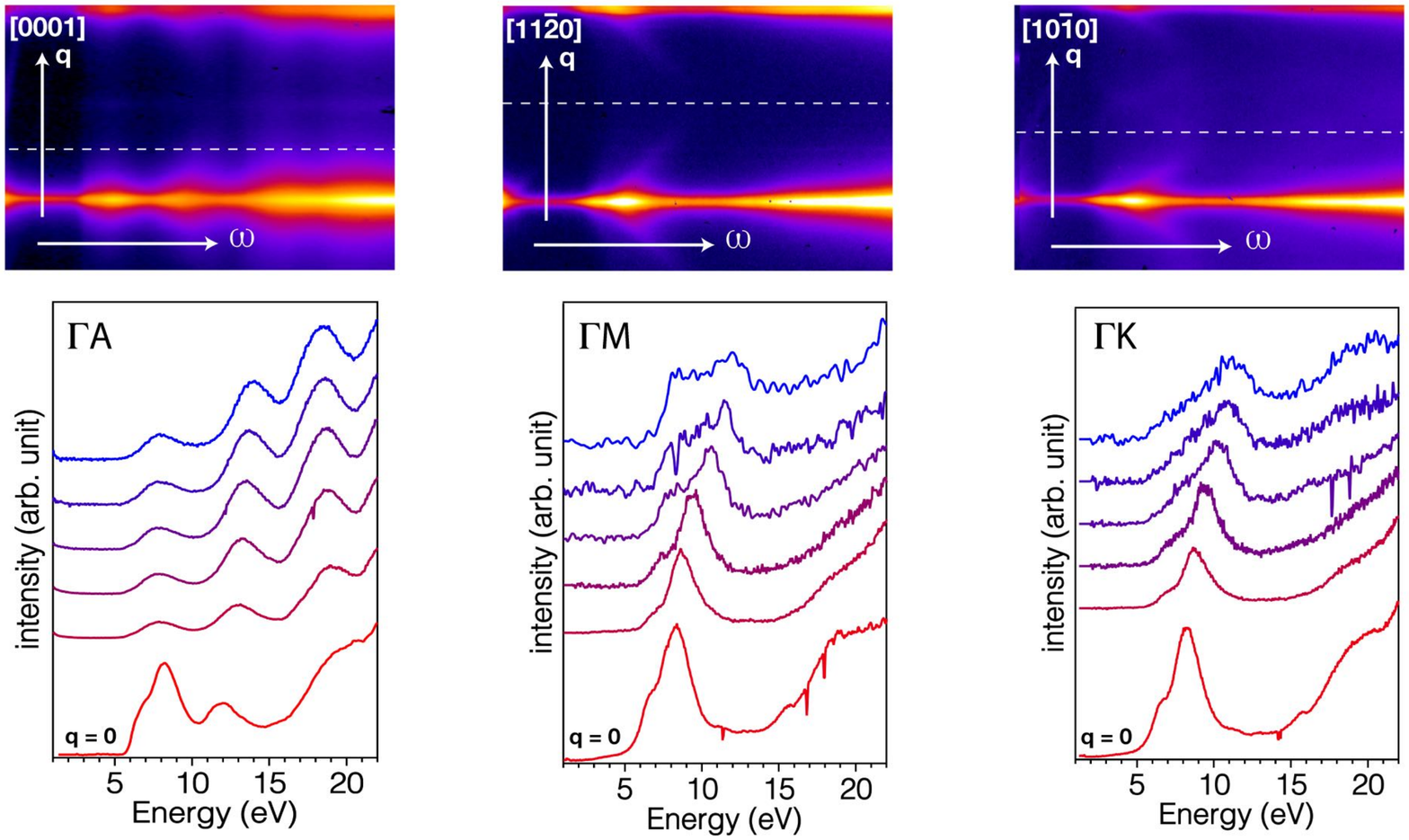} 
\caption{Top panels: $\omega-\bm{q}$ plots as measured (dotted lines indicate the Brillouin zone boundaries). Bottom panels: Loss function spectra along high-symmetry directions. Spectra at $q=0$ are shown separately, while the others are distributed every 20\% of the Brillouin zone.}
\label{lowlossomega}
\end{center}
\end{figure*}
\begin{figure}[b]
\begin{center}
\includegraphics*[width=8cm]{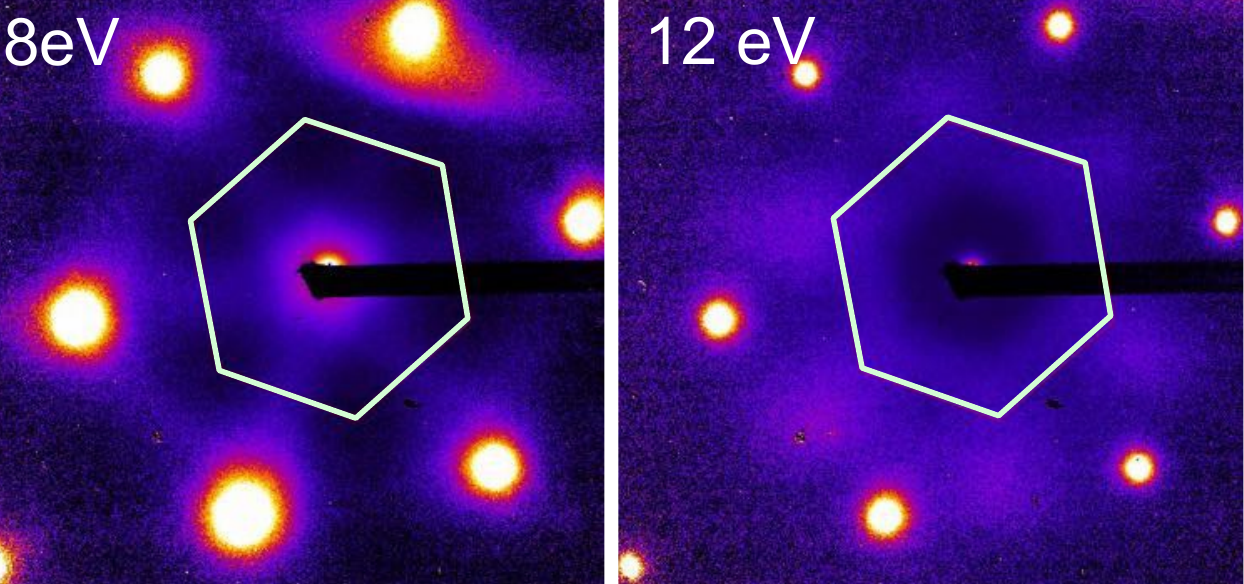} 
\caption{Energy-filtered scattering patterns measured at different energies in the low-loss regime. The Brillouin zone boundary is marked in white.}
\label{scatt}
\end{center}
\end{figure}

The low loss regime is related to the loss function, equal to $-\mbox{Im} [1/\varepsilon(\bm{q},\omega)]$ proportional to $S(\bm{q},\omega)/q^2$, so that the differential scattering cross section is proportional to $- 1/q^2\,  \mbox{Im}[1/\varepsilon(\bm{q},\omega)]$. The peaks of the loss function are frequently associated with plasmons. Two energy ranges are generally distinguished, with a $\pi$ plasmon peak in the  6--8 eV range and a $\sigma +\pi$ peak at about 25 eV for bulk $h$-BN and also for graphite,\cite{Tarrio1989,Moreau2003,Marinopoulos2004,Nicholls2012,Pan2012} the position and the intensity of the latter peak being strongly dependent on the number of sheets in thin samples. The position of some structures can also be associated to specific interband transitions, particularly if they are correlated to the behaviour of $\varepsilon(\bm{q},\omega)$ itself through Kramers-Kronig analyses,\cite{Tarrio1989} but some controversy has appeared recently between these two interpretations and concerning the nature of the observed signals in 2D systems such as graphene.\cite{Nelson2014,Novko2015,Liou2015,Nazarov2015} It is not obvious actually to derive well-defined dispersion relations and to decide between the two possibilities. In most cases the excitations have a mixed character reinforced by the fact that local field and many-body effects are important, so that the discussion has a somewhat semantic character. Nevertheless accurate calculations based on the Bethe-Salpeter equation are now available and recently have been used successfully to analyse NRIXS experiments,~\cite{Galambosi2011, Fugallo2015} notably in relation with specific excitonic peaks arising at $\bm{q}$ outside the first Brillouin zone.

The capabilities of our approach in the low-loss regime are well exemplified by Fig.~\ref{lowlossomega} and Fig.~\ref{scatt}, that report images produced by the two recording techniques of the spectroscopic set-up. In Fig.~\ref{lowlossomega} we present $\omega-\bm{q}$ maps (top panels) as well as the corresponding loss function along the $\Gamma A$, $\Gamma M$ and $\Gamma K$ directions in the range 0--25~eV (bottom panels). Along $\Gamma K$ the low-energy peak moves upwards, from about 8~eV to about 12~eV whereas along $\Gamma M$ this peak splits when $\bm{q}$ approaches the Brillouin zone boundary at $M$. The two spectra instead look similar at small $\bm{q}$ and coincide at $\bm{q}\approx 0$. Along $\Gamma A$ the peaks do not disperse significantly as a manifestation of the weak inter-planara interaction. The fact that the $\Gamma A$-spectrum  differs significantly from the other two is because of the anisotropy of the dielectric function $\varepsilon_\parallel(\omega) \neq \varepsilon_\perp(\omega)$. 
As a complementary piece of information, we also have access to scattering patterns at fixed energy in both basal and prismatic orientations. As an instructive example, Fig.~\ref{scatt} shows two patterns taken by averaging the signal in the basal plane within 1~eV around 8~eV and 12~eV. Although multiple scattering effects spoil the signal outside the Brillouin zone,\cite{Weissker2010} inside it the diffuse intensity, which is the relevant quantity here, is well detectable. These maps clearly show the differences in the dispersion of the diffuse intensity observed from the $\omega-\bm{q}$ maps along $\Gamma K$  and $\Gamma K$ at about 8~eV and 12~eV. 

 In the following, we illustrate the novelties of our method applied to the low-loss regime by discussing its complementarity to X-ray spectroscopy.~\cite{Galambosi2011} Then we will compare \textit{ab initio} calculations against our data with the intent to assess the level of approximation required for an accurate description.

\subsubsection{Comparing EEL and NRIX spectroscopies}

The small-$q$ regime is particularly appealing to discuss the strong points of EELS with respect to NRIXS. It is indeed recalled here that the scattering cross section is proportional to $q^2 \,\mbox{Im}[-1/\varepsilon(\bm{q},\omega)]$ in the case of NRIXS, and proportional to $(1/q^2)  \,\mbox{Im}[-1/\varepsilon(\bm{q},\omega)]$ in the case of EELS. This makes EELS particularly suited for probing small exchanged momenta. This complementarity is evident when comparing our data (Fig.~\ref{lowlossomega}) to NRIXS ones.~\cite{Galambosi2011}  In Fig.~\ref{lowlossomega}, the signal starts becoming noisy at $q\approx1.0$~\AA$^{-1}$ (around 60\% of the $\Gamma K$ and 80\% of $\Gamma M$); conversely in Fig.1 of reference [\onlinecite{Galambosi2011}], the signal is extremely weak up to 0.6~\AA$^{-1}$, which is even beyond the zone border along $\Gamma A$. Being intrinsically very sensitive in the small-$q$ regime, our EELS method can bridge the gap between optical measurements (very precise but limited to $q\rightarrow 0$) and the X-ray experiments (sensitive at large $\bm{q}$).
It is hence a powerful and versatile tool to make accurate investigations of the dielectric properties inside the Brillouin zone, and notably in the vicinity of $\bm{q}=0$ where excitonic effects exhibit peculiar characteristics in 2D-materials and thin films.~\cite{Cudazzo2016,Latini2015}

However, in the optical limit results has to be analysed carefully. When comparing $\Gamma K$ and $\Gamma M$ directions at $\bm{q}\approx 0$, it is clear that the scattered intensity becomes isotropic in the basal plane (see bottom panels of Fig.~\ref{lowlossomega}), as expected from physical considerations and in agreement with theoretical calculations\cite{Cappellini2001,Marinopoulos2004,Guo2005}. In the same limit, the spectrum along $\Gamma A$ differs from the in-plane ones because of the anisotropy of $\varepsilon(\bm{q},\omega)$ but it exhibits an intense structure at 8 eV which is unexpected. In fact most calculations predict a much weaker intensity for structures below 12 eV\cite{Cappellini2001,Marinopoulos2004,Guo2005}. 
In the core losses, the $\bm{q}\to 0$ limit was problematic because of the $q_E$ component. But here $q_E$ can be neglected as it is of the order of 0.005~\AA$^{-1}$ owing to the lower energy loss. In this case actually the problem comes from the width $\Delta q_y$ of the slit used in the $\omega-\bm{q}$ mode which is about 0.20~\AA$^{-1}$. When collecting data along $\Gamma M$ and $\Gamma K$, the slit lays parallel to the basal plane, where $h$-BN is isotropic at $\bm{q} \to 0$. Instead when measuring along the $\Gamma A$ direction, $q_x$ is parallel to $z$ while $q_y$ still lays parallel to the basal plane. This leads to a mixture of $\varepsilon_\parallel$ and $\varepsilon_\perp$, the latter being predominant. This explains why the 8~eV structure in the $\bm{q}\approx 0$ $\Gamma A$-spectrum looks so similar to the equivalent peak in the basal-plane spectra and is instead washed out at higher $\bm{q}$.

\begin{figure}
\begin{center}
\includegraphics*[width=8cm]{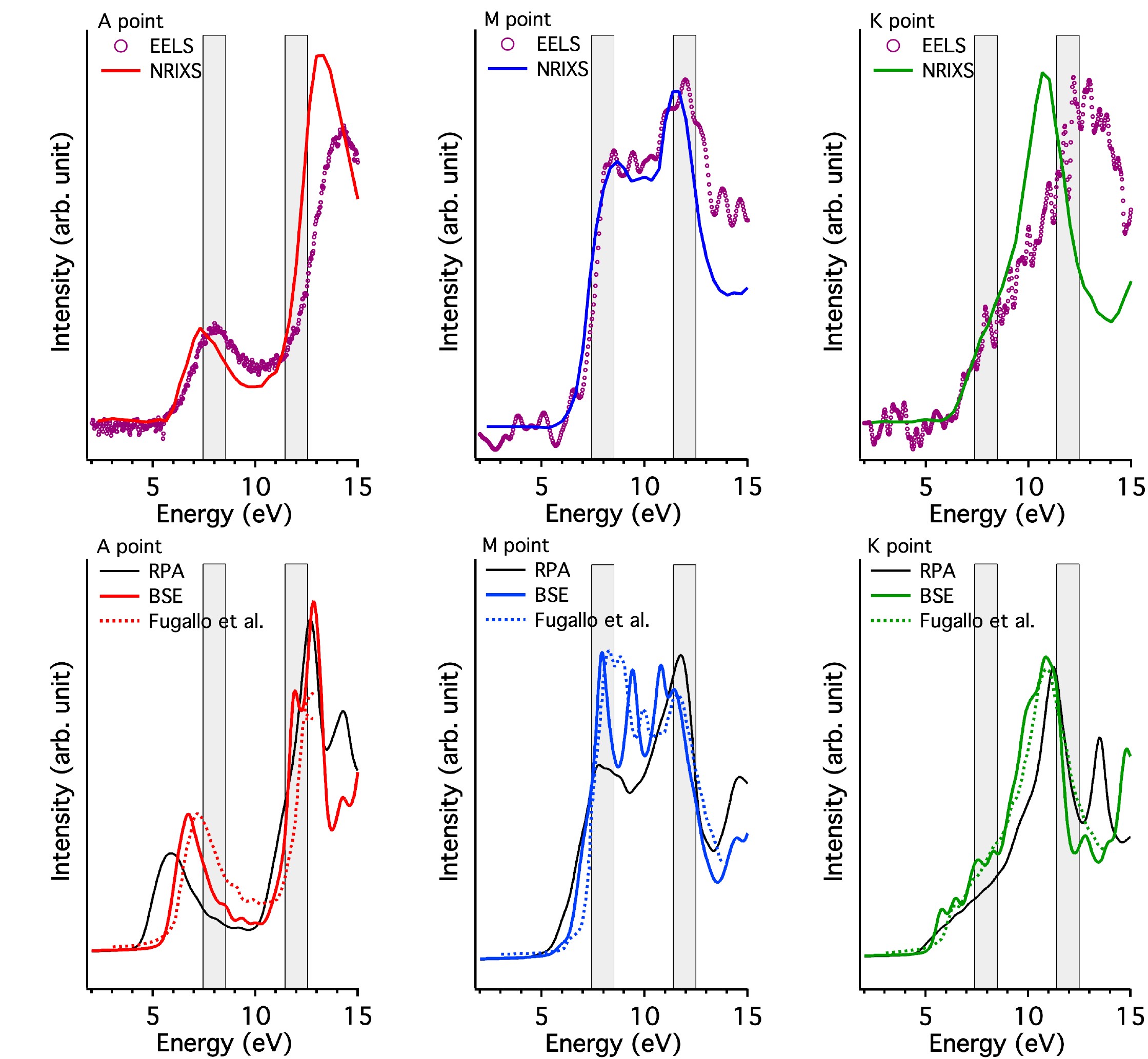} 
\caption{Top panels: Comparison between our EELS and NRIXS~\cite{Galambosi2011} measures. Bottom panels: Comparison between theoretical curves (ours and Fugallo's\cite{Fugallo2015}). Our theoretical spectra have been convoluted with a Gaussian of width~0.2 eV.}
\label{edge}
\end{center}
\end{figure}

We now point out that the ``low-$q$" region is actually wide enough to explore the entire first Brillouin zone. In the top panels of Fig.~\ref{edge} we report the comparison between our data and NRIXS ones\cite{Galambosi2011} at the high-symmetry points $A$, $M$ and $K$ located at the zone boundary. In these points both techniques have reasonably high accuracy. Grey-shaded strips delimit the energy intervals for the averages done in obtaining the diffraction maps of Fig.~\ref{scatt}. The good agreement between the two techniques demonstrate that the EELS give accurate results for $\bm{q}$ as large as the zone border. Together with the considerations above, this shows that the EELS ensures high-quality data inside the whole Brillouin zone.

\subsubsection{Comparison with theoretical calculations}

As discussed by Galambosi \textit{et al.}~\cite{Galambosi2011} and by Fugallo \textit{et al.}~\cite{Fugallo2015}, several  theoretical ingredients are necessary to account for all the details of inelastic scattering experiments. This is especially true for excitonic features that can be correctly simulated only by going beyond the independent-particle approach (RPA) and solving instead the Bethe-Salpeter equation (BSE). In the works cited above, the authors recurred to BSE to investigate the origin of specific peaks, in particular for momenta outside the first Brillouin zone. Though, BSE is computationally very demanding with respect to RPA. The question we want to answer in this section is then ``\emph{Is the BSE accuracy  indispensable to describe our EELS data?}"   

All calculations have been carried out with the code GPAW\cite{Mortensen2005}. Structural parameters are $a$=2.50 \AA\,and $c$=6.5 \AA, in agreement with those obtained from measured diffraction patterns. The ground state density has been obtained sampling the Brillouin zone with a ($6 \times 6 \times 2$) $\Gamma$-centred k-point grid and including plane waves up to 900~eV. The Perdew-Burke-Ernzerhof approximation has been adopted for the exchange-correlation potential\cite{Perdew1996}. The RPA loss function has been computed in all q-points of a ($24 \times 24 \times 8$) $\Gamma$-centred grid, including 20 bands and with a cutoff of 60~eV. The diffraction patterns at energy $E$ have been obtained by first averaging the computed spectra in the range $E \pm 0.5$~eV and then interpolating the result on a sufficiently dense mesh ($50 \time 50$ points). The BSE has been solved for six valence bands and eight conduction bands on a ($12 \times 12 \times 4(8)$) $\Gamma$-centred q-point grid for $\mathbf{q}$ in-plane (out-of-plane). A scissor operator of 1.73~eV (derived from the average GW correction across the gap) has been applied to the PBE energies. A cutoff of 60~eV and 20 bands have been included to converge the dielectric constant entering in the direct term of the excitonic Hamiltonian.

In the bottom panel of Fig.~\ref{edge} we report our RPA and BSE curves together with data extracted from the work by Fugallo and coworkers\cite{Fugallo2015} used as a validation benchmark. The main differences between RPA and BSE spectra are (i) a redistribution of the spectral weight, notably in the first peaks at $\bm{q}=M$, and (ii) a shift of the low-energy peak at $\bm{q}=A$. These excitonic effects have been already discussed in literature\cite{Galambosi2011,Fugallo2015} and further details can be found in the Supplemental Material. Indeed to account for these relatively tiny effects BSE is unavoidable, but what we want to stress here is that at $\bm{q}=M$ and $\bm{q}=K$ not only the main structures are correctly reproduced already at the RPA level, as expected, but they also fall in the right position. Moreover, this is true in the whole basal plane (cfr. Fig.6 of Supplemental Material).

The physical reason is that on the basal plane the quasiparticle normalisation due to e-e scattering (here approximated by the scissor operator) is almost entirely cancelled by the e-h attraction. From a practical point of view this means that, as long as $\bm{q}$ lays on the basal-plane, the RPA is good enough to describe the dispersion of the loss function and it can be used successfully to simulate both acquisition methods. Instead along $\Gamma A$, the anisotropy of the electronic screening spoils this mutual cancellation, leading to a misalignment of the first peak. This is clearly shown in the left panels of Fig.~\ref{edge}, where the measured data exhibit a local maximum at 8~eV, whereas the RPA spectrum is almost vanishing. Both data instead overlap pretty well at $\sim$12~eV. As a consequence, the loss function with $\bm{q}\parallel \Gamma A$ can not be computed at the RPA level in a large energy range with the right alignment of all peaks. In particular this is a problem when simulating diffraction patterns (eg. in the $\Gamma A M$ plane) since a single plot includes perpendicular $\bm{q}$ (correctly aligned), parallel $\bm{q}$ (wrongly aligned), and all momenta in between. The right alignment in all directions can be surely ensured by BSE, but the heavy computational cost of the method hinders the applicability to the simulation of diffraction patterns. Moreover, the energy average carried out would wash out most of the weight redistribution, which makes the use of BSE quite disproportionate.

\begin{figure}
\begin{center}
\includegraphics*[width=8cm]{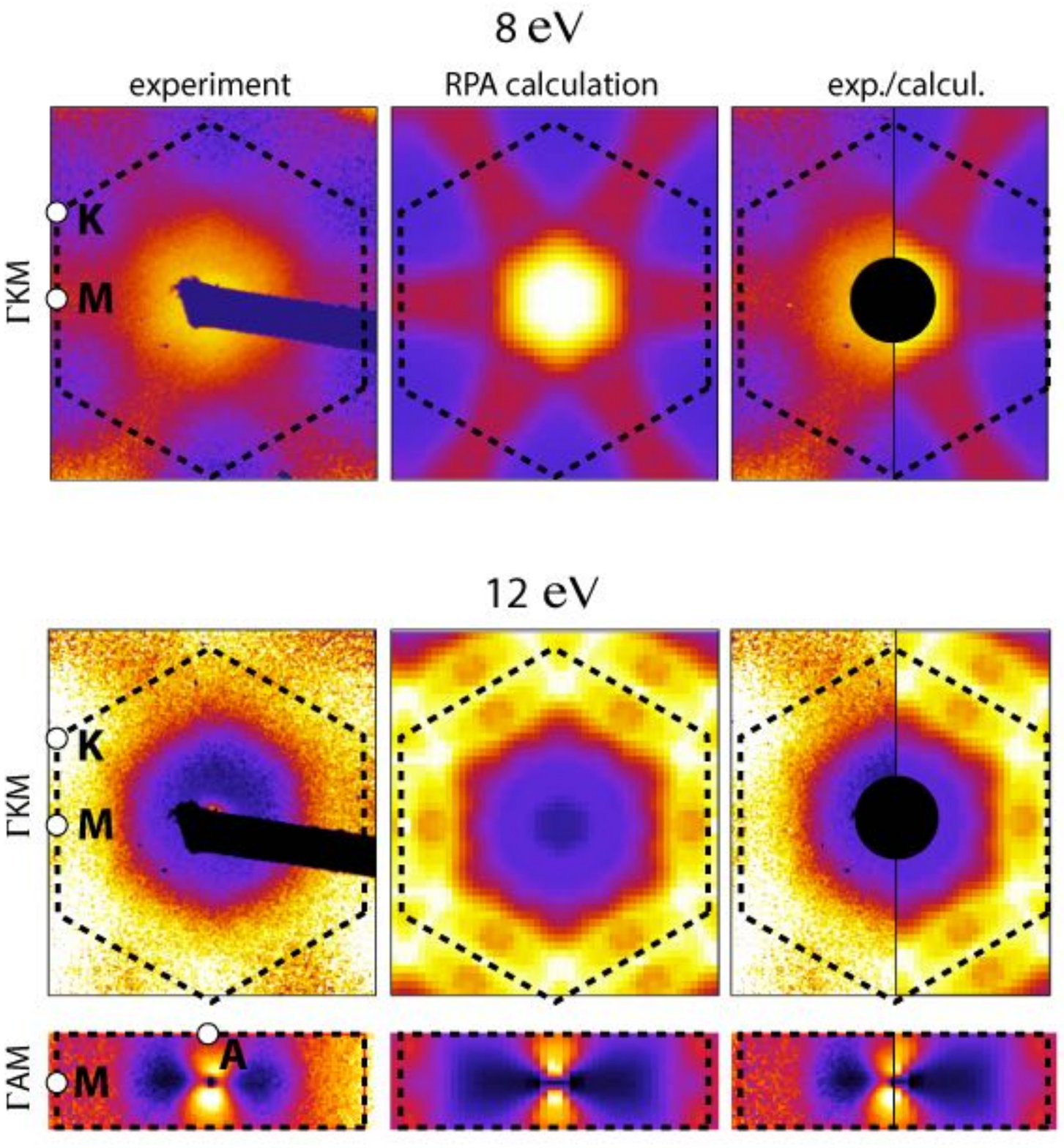} 
\caption{Experimental and calculated loss functions (scattered intensities multiplied by $q^2$)
at 8 eV for the $\Gamma KM$ plane and at 12 eV for the $\Gamma KM$ and $\Gamma AM$ planes. Dashed lines delimitate the Brillouin zone.}\label{comp}
\end{center}
\end{figure}

In order to display the quality of the RPA, in Figure~\ref{comp} we report simulated and measured energy-filtered dispersion patterns of the loss function at energies 8~eV and 12~eV in the basal plane (top and central panels). At low energy (8~eV), the patterns in the $\Gamma KM$ plane are characterized by an intensity concentrated at the origin, with diffuse arms pointing along the $\Gamma M$ directions, whereas at higher energy the intensity is higher close to the Brillouin zone boundary, with diffuse arms along the $\Gamma K$ directions. This is consistent with a simple analysis in terms of $\pi-\pi^*$ excitations in this  energy regime. At low energy the transitions are mainly direct transitions ($\bm{q}\simeq 0$), whereas they are indirect at higher energy\cite{Galambosi2011}. At the bottom of Fig.~\ref{comp}, we show similar maps at 12~eV in the $\Gamma A M$ plane which contains therefore the $\Gamma A$ direction. This has been possible because at this energy the signal is accidentally well aligned in all directions. As expected from the spectra in Fig.~\ref{lowlossomega} the intensity at low $q$ is maximum in the $\Gamma A$ direction.

\section{Conclusion}

In summary we demonstrate that momentum-resolved EELS allows us to obtain accurate information on the electronic excitation spectra for core losses as well as for low losses. We illustrate this by treating the case of $h$--BN for which NRIXS data are available. The energy filtered diffraction patterns provide a global view of anisotropy effects in $\bm{q}$ space, whereas the  $\omega-\bm{q}$ plots allow us to map the symmetries of the losses as a function of the energy transferred to the material. The case of core losses related to the excitonic $\sigma-\pi^*$ transitions at the boron $K$ edge has been shown to be particularly spectacular. In the case of low losses, our results confirm those of inelastic x-ray scattering experiments allowing us to point out some advantages specific of our method.

EELS is an efficient technique complementary to other inelastic scattering tools such as NRIXS, despite their comparable energy resolution of about 100-200 meV. Indeed their accuracy in $\bm{q}$-space is different. As EELS performs much better at low $q$, it opens the way to make contact with optical measurements. In particular, in 2D materials and heterostructures  excitonic effects are important and present peculiar characteristics close to $\bm{q}=0$\cite{Cudazzo2016,Latini2015} , it is therefore highly desirable to have a tool adapted for measuring the dispersion of the excitonic levels in the low $\bm{q}$ regime. Moreover EELS within an electron microscope has the nonnegligible advantage of permitting fast and local experiments at the nanoscale.

Finally, with the support of \textit{ab initio} calculations at the RPA and BSE level, we pointed out that in $h$--BN e-e and e-h effects almost cancel out in excitations with $\bm{q}$ parallel to the layers, while the former dominates for $\bm{q} \parallel \Gamma A$. The practical consequence is that, as long as the exchanged momentum lays in-plane, RPA calculations describe well energy-loss spectra and diffraction patterns in a pretty large energy range. Instead, when excitations perpendicular to the planes are involved, one has to rely on BSE calculations to correctly align all the peaks. This is particularly problematic when simulating diffraction patterns because of the high computational cost of these calculations.

\begin{acknowledgments}
T. Taniguchi and K. Watanabe from NIMS are warmly acknowledged for providing a reference HPHT crystal. The authors want to thank David Troadec from IEMN for the FIB samples preparation, and Philip Wachsmuth, Gerd Benner and Ute Kaiser for very useful discussions on $\omega-\bm{q}$ maps. 
Hakim  Amara is acknowledged for many fruitful discussions, Giorgia Fugallo for the careful comparison with her theoretical results, and Etienne Gaufr\'es for a careful reading of the manuscript. The research leading to these results has received funding from the European Union H2020 Programme under grant agreement no. 696656 GrapheneCore1. We acknowledge funding by the French National Research Agency through Project No. ANR-14-CE08-0018. 
\end{acknowledgments}


%

\newpage

\begin{widetext}

\section*{Supplementary Material of : Angular resolved electron energy loss spectroscopy in hexagonal boron nitride}

\renewcommand\thefigure{S\arabic{figure}}
\setcounter{figure}{0}

\subsection{HRTEM of FIB slabs}
\begin{figure}[!ht]
\begin{center}
\includegraphics*[width=12cm]{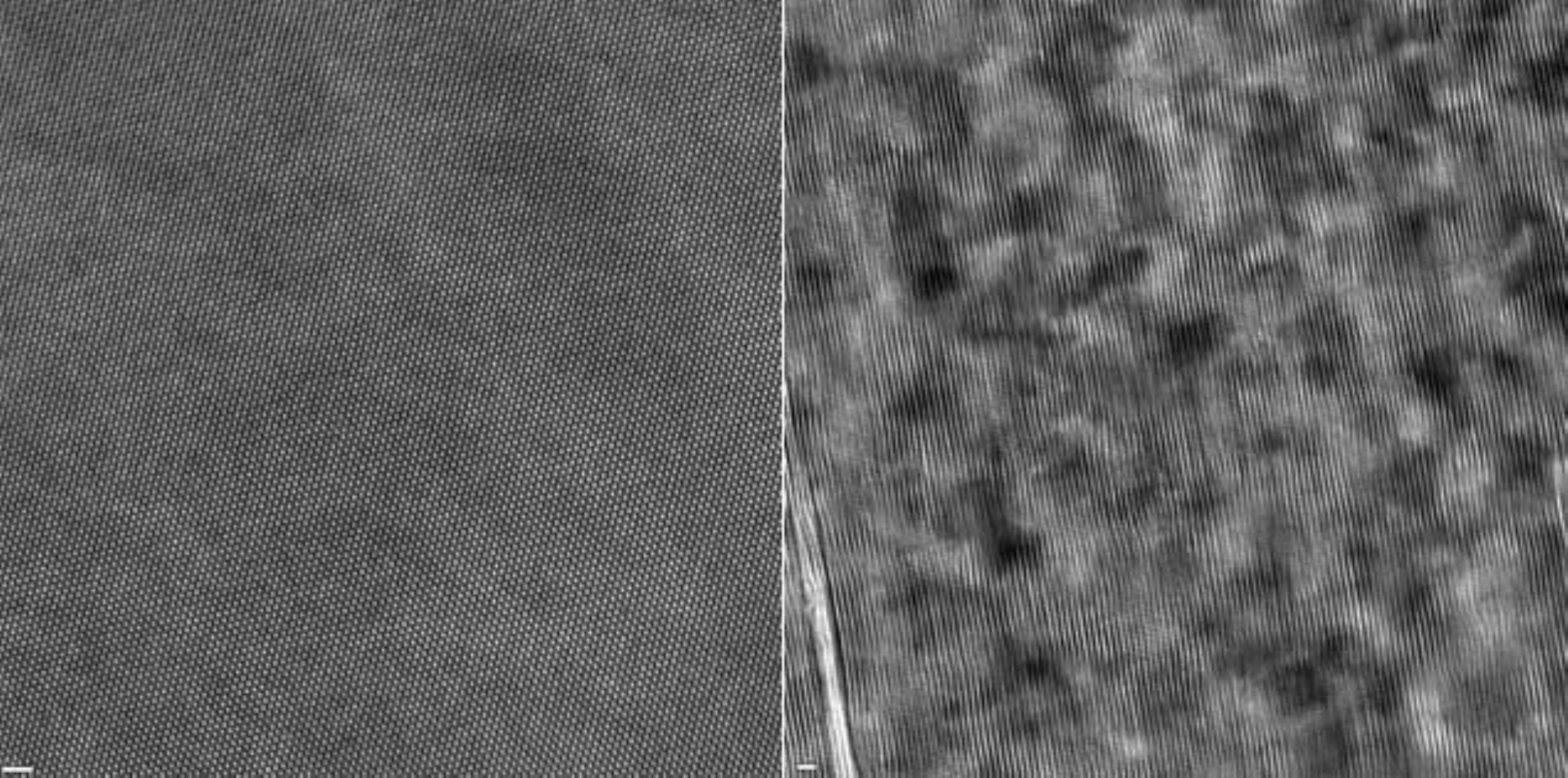} 
\caption{ High Resolution TEM image of samples prepared by FIB. Sample orientations correspond to  $(0001)$ zone axis (left) and $(10\bar{1}0)$ zone axis (right).}
\end{center}
\end{figure}
\newpage

\subsection{$\omega-q$ map at the nitrogen $K$ edge  [0001] direction}
\begin{figure}[!ht]
\begin{center}
\includegraphics*[width=8cm]{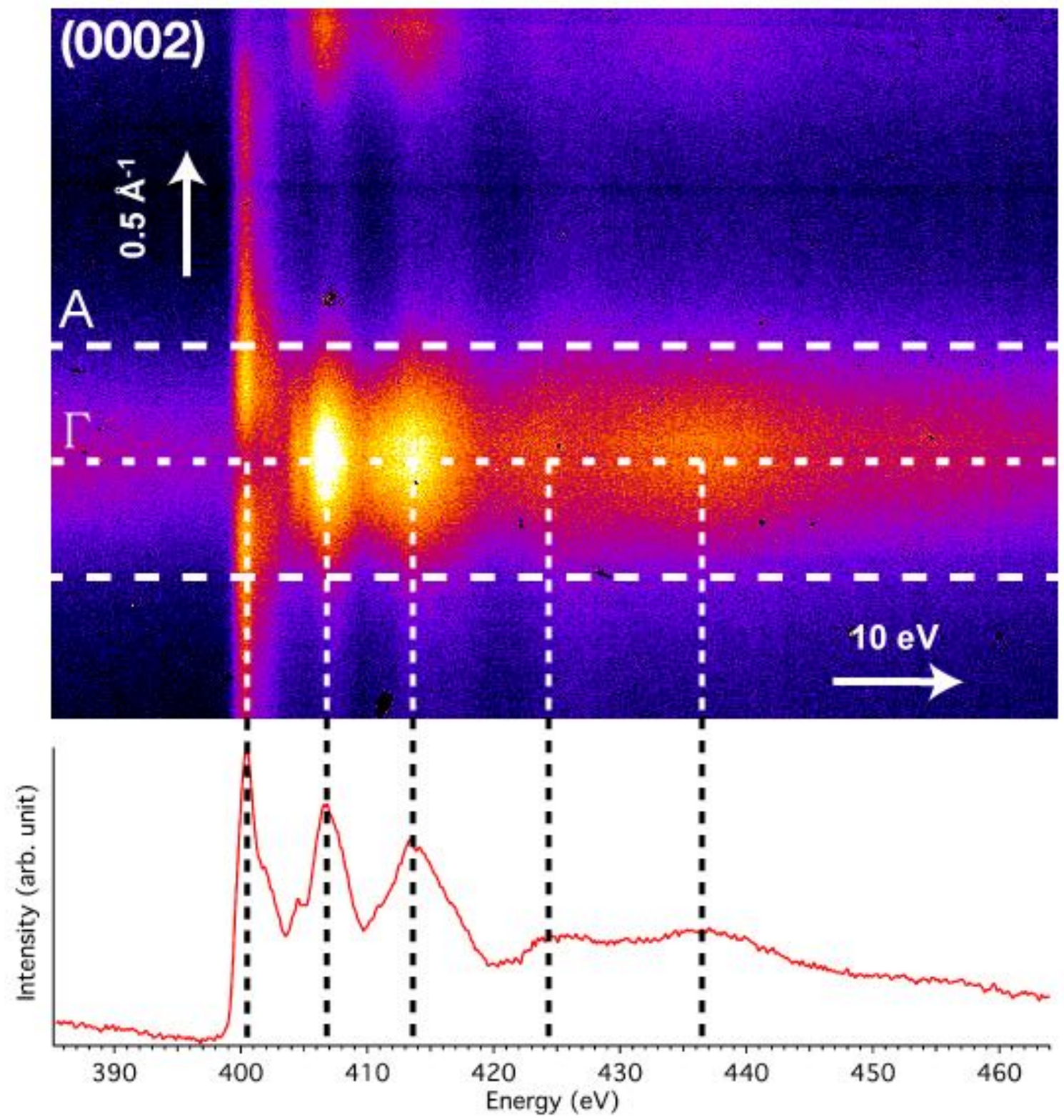} 
\caption{$\omega-\bm{q}$ image recorded in the $\Gamma A$ direction close to the nitrogen K-edge  (400 eV). Dashed lines delimitate the Brillouin zone. Bottom : related EELS spectrum integrated over the whole Brillouin zone. The dotted line indicates the corresponding Brillouin zone section and significant edge structures. }
\end{center}
\end{figure}
\newpage

\subsection{$\omega-\bm{q}$ geometry versus crystal orientation}
Loss spectra are very dependent on the orientation of the sample with respect to the electron beam. Details concerning the geometry used in the experiments are given here.
\begin{figure}[!ht]
\begin{center}
\includegraphics*[width=15cm]{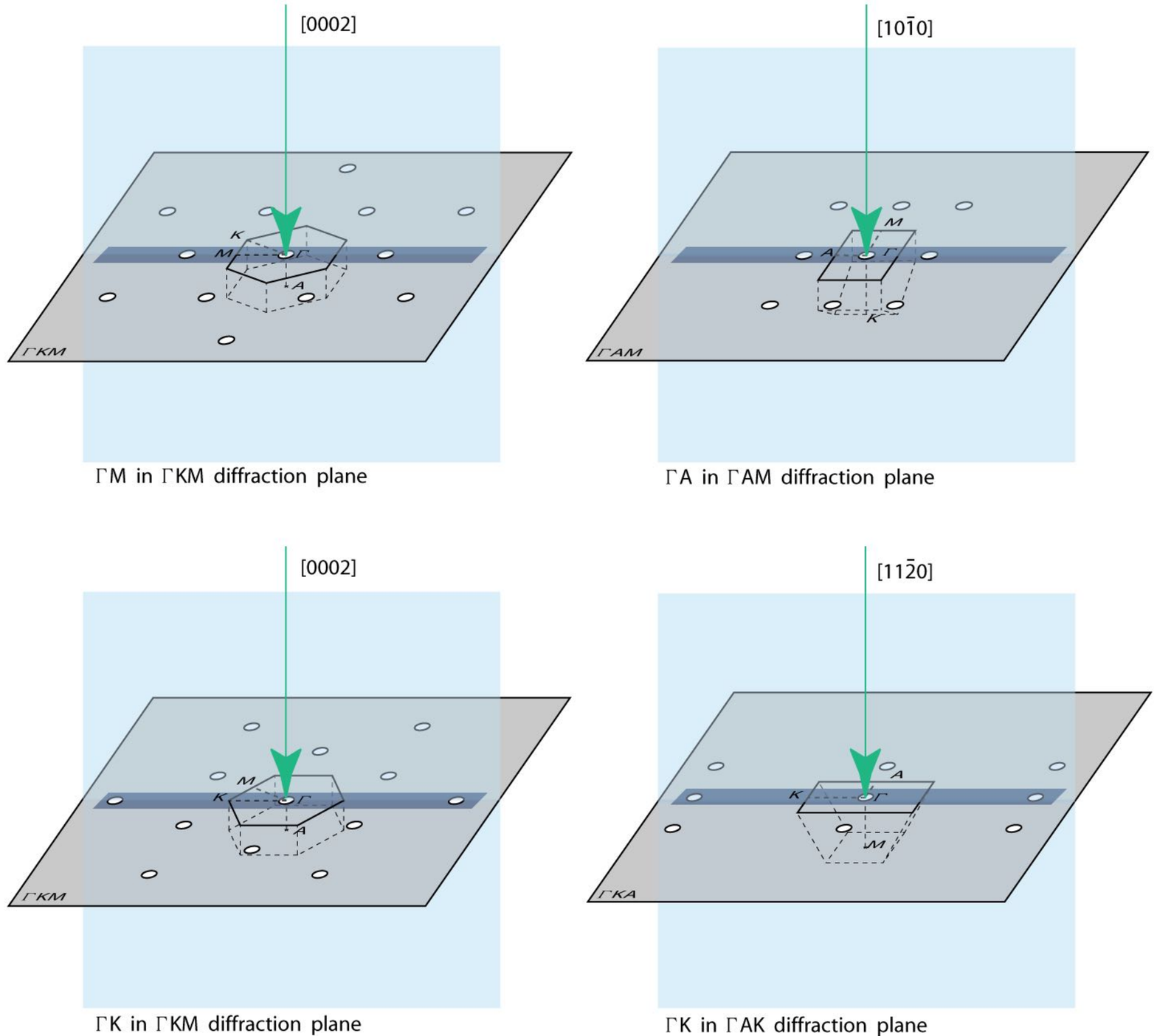} 
\caption{Different instrumental geometries related to the crystal orientation (zone axis) and the direction of the slit at the entrance of the $\Omega$ filter. The spectra in the $\Gamma M$ and $\Gamma K$ directions shown in the main text correspond to the $[0002]$ zone axis (left). The spectrum in the $\Gamma A$ direction corresponds to the $[10\bar{1}0]$ axis.}
\label{geometry}
\end{center}
\end{figure}
As can be seen in Fig.\  \ref{geometry}, the zone axis is not sufficient to describe the geometry and one has also to precise the orientation of the slit used in  $\omega-\bm{q}$ maps. The diffraction pattern of the sample as well as the Brillouin zone are shown in the Fig.  \ref{geometry}. One can see  that, depending on the geometry, it is possible to probe a direction with different components in and out of the diffraction plane. In the main text the spectra along $\Gamma M$ and $\Gamma K$ correspond to the $[0002]$ zone axis (left). The spectrum along $\Gamma A$ corresponds to the $[10\bar{1}0]$ zone axis. Fig.  \ref{lloss_new} shows results obtained with  geometries different from those  in the main text. The results are identical for $q \neq 0$ whereas there are differences in the $q = 0$ spectrum because of the different components (\textit{i.e.} Brillouin zone direction) mixed at $ q = 0$.

\begin{figure}[!ht]
\begin{center}
\includegraphics*[width=10cm]{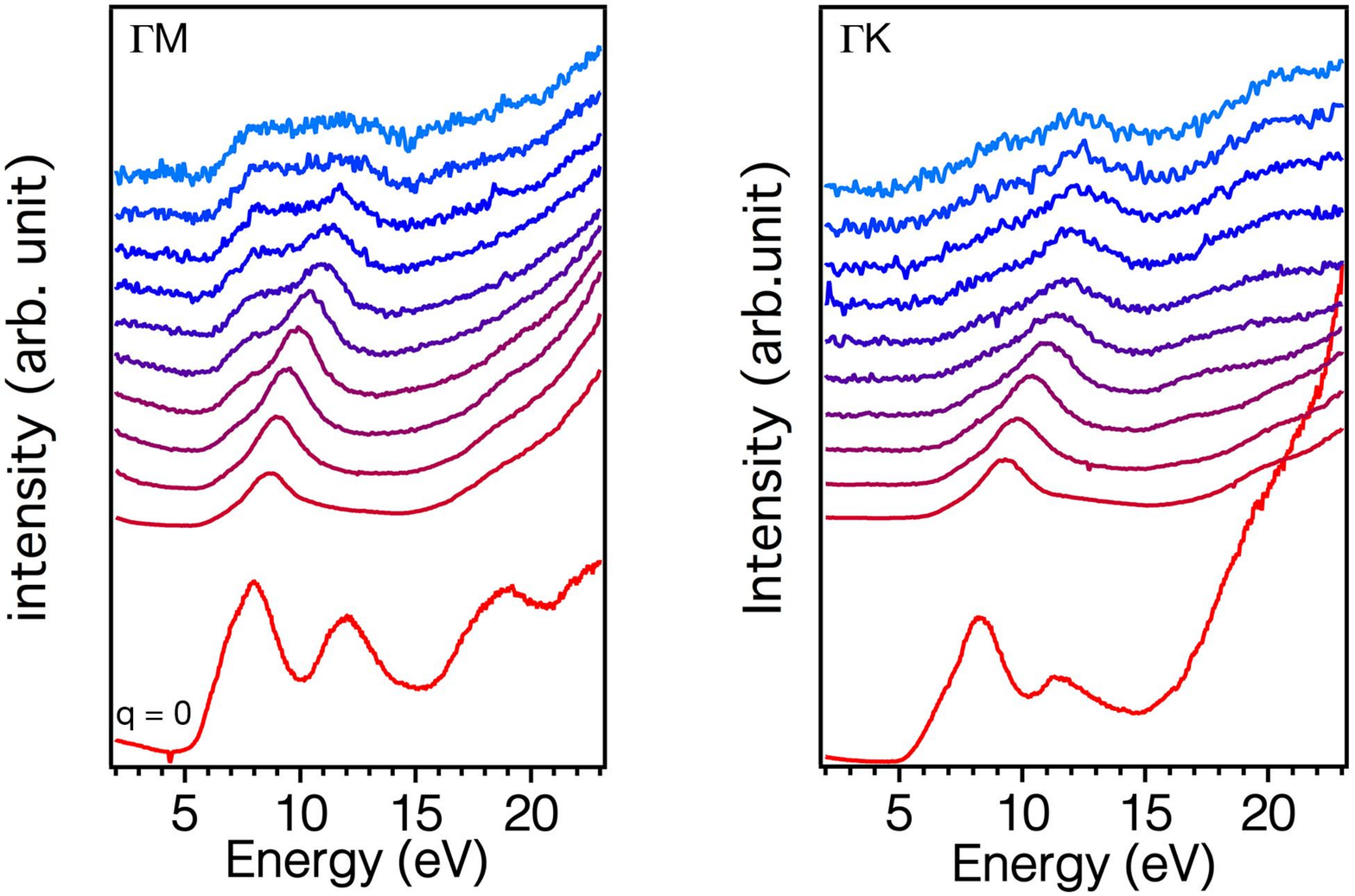} 
\caption{$\omega-q$ plot and EEL spectra multiplied by $q^2$ (for $q \neq 0$) along the $\Gamma M$ and $\Gamma K$ directions. Left : zone axis is $[10\bar{1}0]$, diffraction plane is $(\Gamma MA)$, slit is along $[\Gamma M]$. Right : zone axis is $[11\bar{2}0]$, diffraction plane is $(\Gamma KA)$, slit is along $[\Gamma K]$.}
\label{lloss_new}
\end{center}
\end{figure}

\newpage

\subsection{Benchmark EELS, NRIXS from Galambosi et al. and calculations from Fugallo et al.}
\begin{figure}[!ht]
\begin{center}
\includegraphics*[width=12cm]{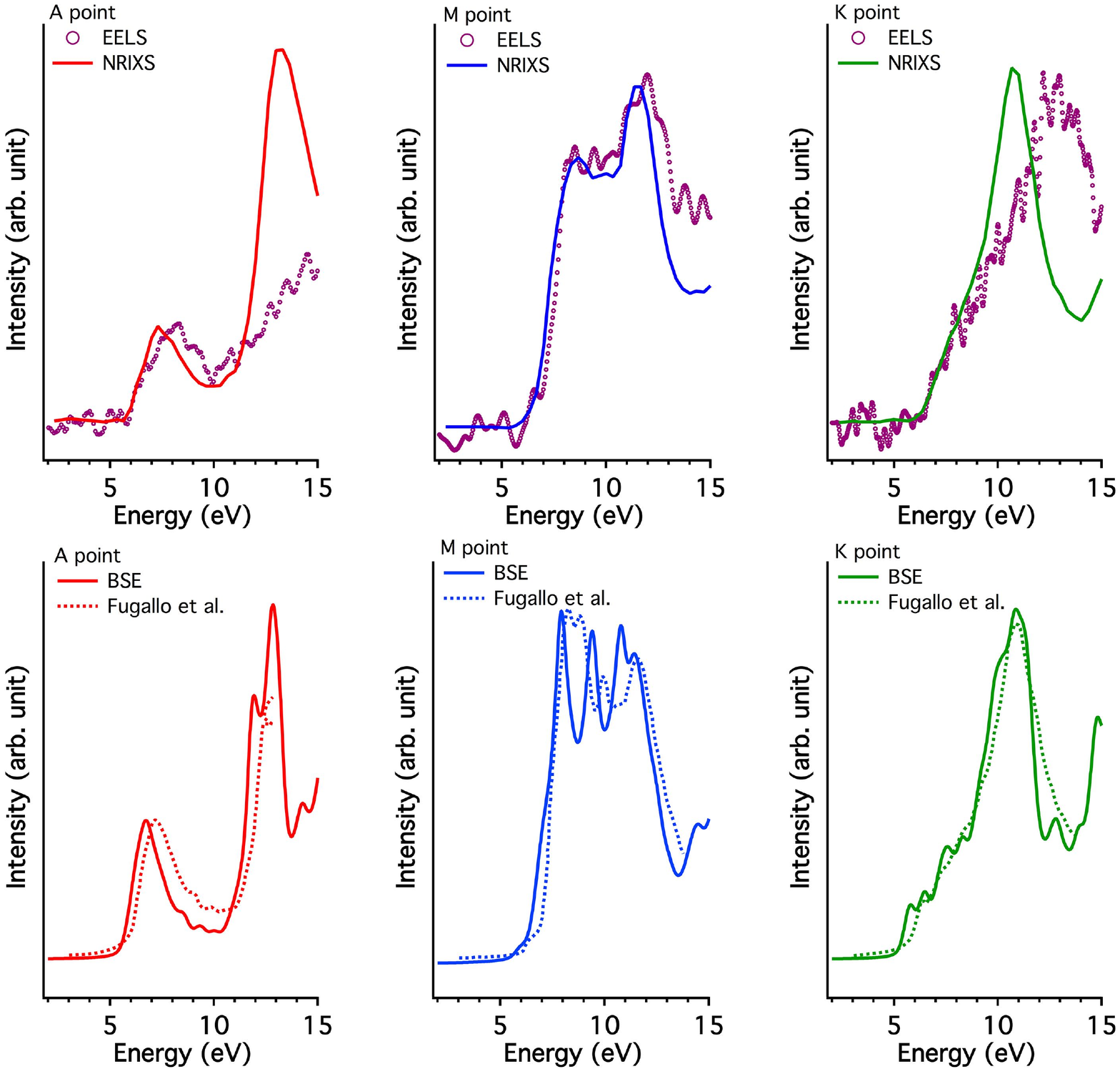} 
\caption{Top: comparison between the present work (EELS) and NRIXS data extracted from Galambosi et al, Phys. Rev. B \textbf{83}, 081413 (2011) at A, M and K points in the Brillouin zone. Bottom: comparison between calculations performed in this work and those extracted from Fugallo et al, Phys. Rev. B \textbf{92}, 165122 (2015).}
\end{center}
\end{figure}

\newpage

\subsection{Theoretical simulations and discussions}

The dielectric matrix $\epsilon_{GG'}(\mathbf{q},\omega)$, has been computed \emph{ab-initio} in the random phase approximation (RPA) and by solving the Bethe-Salpeter equation (BSE). For the latter calculations, the quasiparticle corrections have been approximated by a scissor operator of 1.73~eV (value derived from the average GW correction of the HOMO-LUMO gap). More details on the computational parameters can be found on the main text.

In Fig.\ \ref{fig:eels-lines} we report the loss function extracted from $\omega-\mathbf{q}$ maps along the high-symmetry lines $\Gamma M$, $\Gamma K$ and $\Gamma A$ (dots) together with the dispersion of the RPA spectra along the same directions (solid lines). The agreement for in-plane components is very good in reproducing the main structures, not only at the zone borders but also all along the line. This observation provides an even stronger justification to the use of RPA in drawing the maps of Fig.\  7 of the main text.
On the other hand, we see that the same misalignment pointed for $\mathbf{q}=A$ (see main text) is repeated all along the $\Gamma A$ line, the offset being almost constant. Despite the wrong alignment, theory and experiment agree in predicting a basically dispersionless loss function for off-plane momentum transfer, a confirmation of the fact that plasmonic excitations are well confined on hBN planes.

\begin{figure}[!ht]
\centering
\includegraphics[width=12cm]{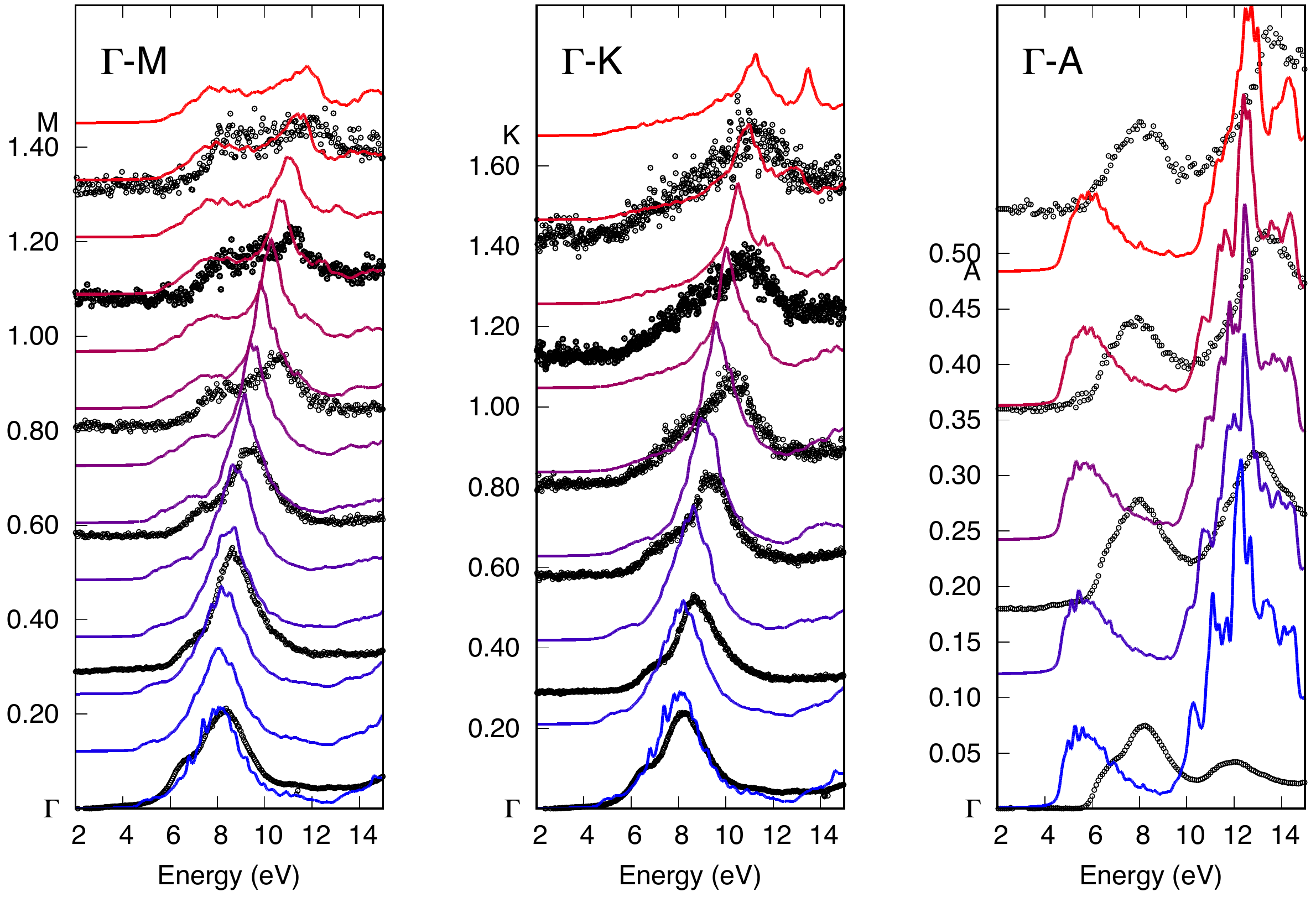}
\caption{RPA and experimental loss function (solid lines and bullets respectively) along high symmetry lines. On $y$ axis units are \AA$^{-1}$.}
\label{fig:eels-lines}
\end{figure}

To identify clearly the excitonic features from other effects, it is instructive to compare the BSE calculation to an independent-particle simulation stemming from the same band structure (we call it SO-RPA). To further reduce the source of discrepancies, we shall use the same q-point grid in both simulations.
The resulting calculations are reported in Fig.\ \ref{fig:excitons}, where solid lines have been computed on the same q-point grid, while the dashed red curves correspond to SO-RPA calculations on a denser q-point grid (respectively $12 \times 12 \times 4 (8) $ k-points in the first case, $24 \times 24 \times 8$ in the second).

\begin{figure}[!ht]
\centering
\includegraphics[width=0.30\textwidth, trim = 20mm 55mm 160mm 20mm, clip]{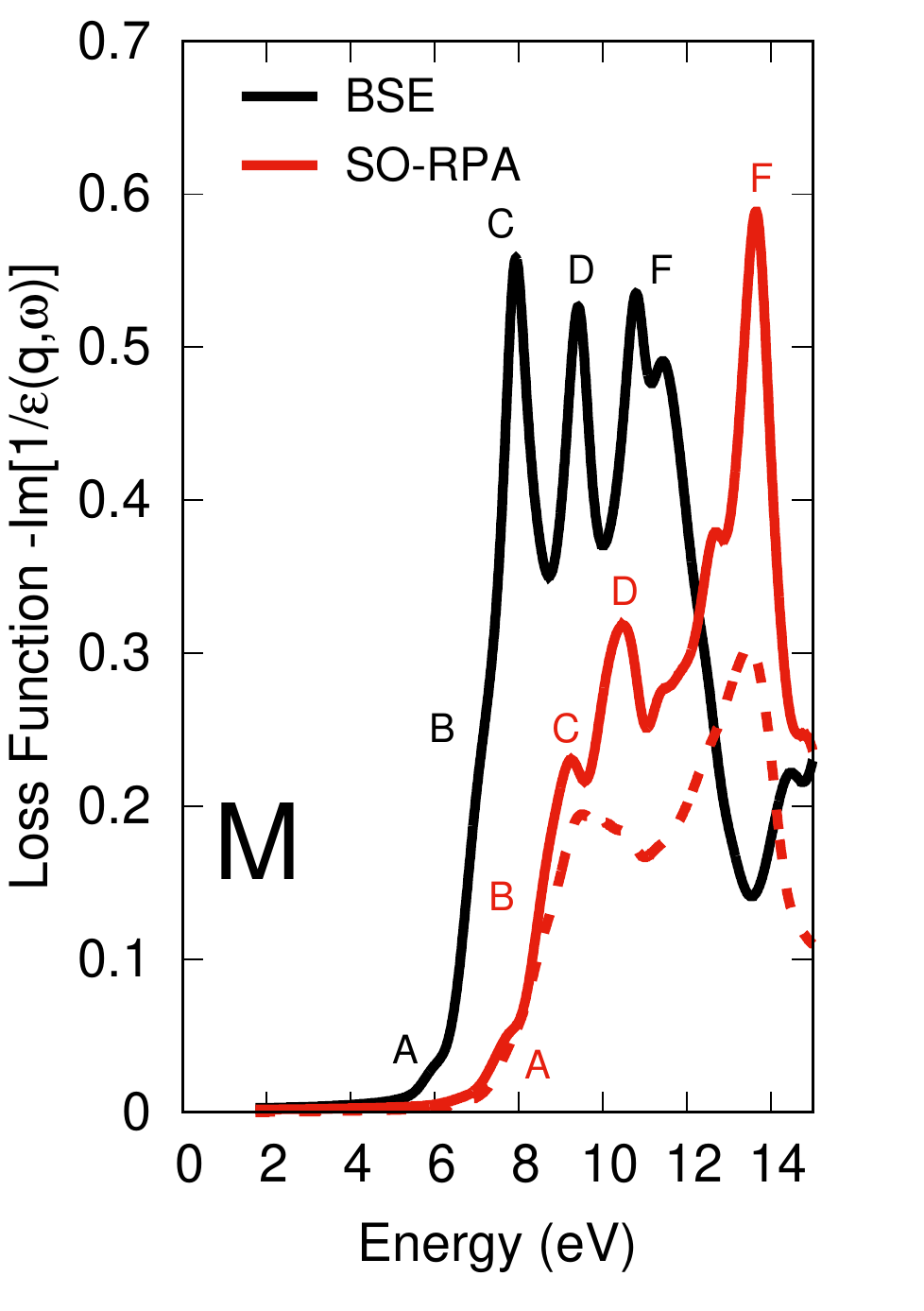}
\includegraphics[width=0.30\textwidth, trim = 20mm 55mm 160mm 20mm, clip]{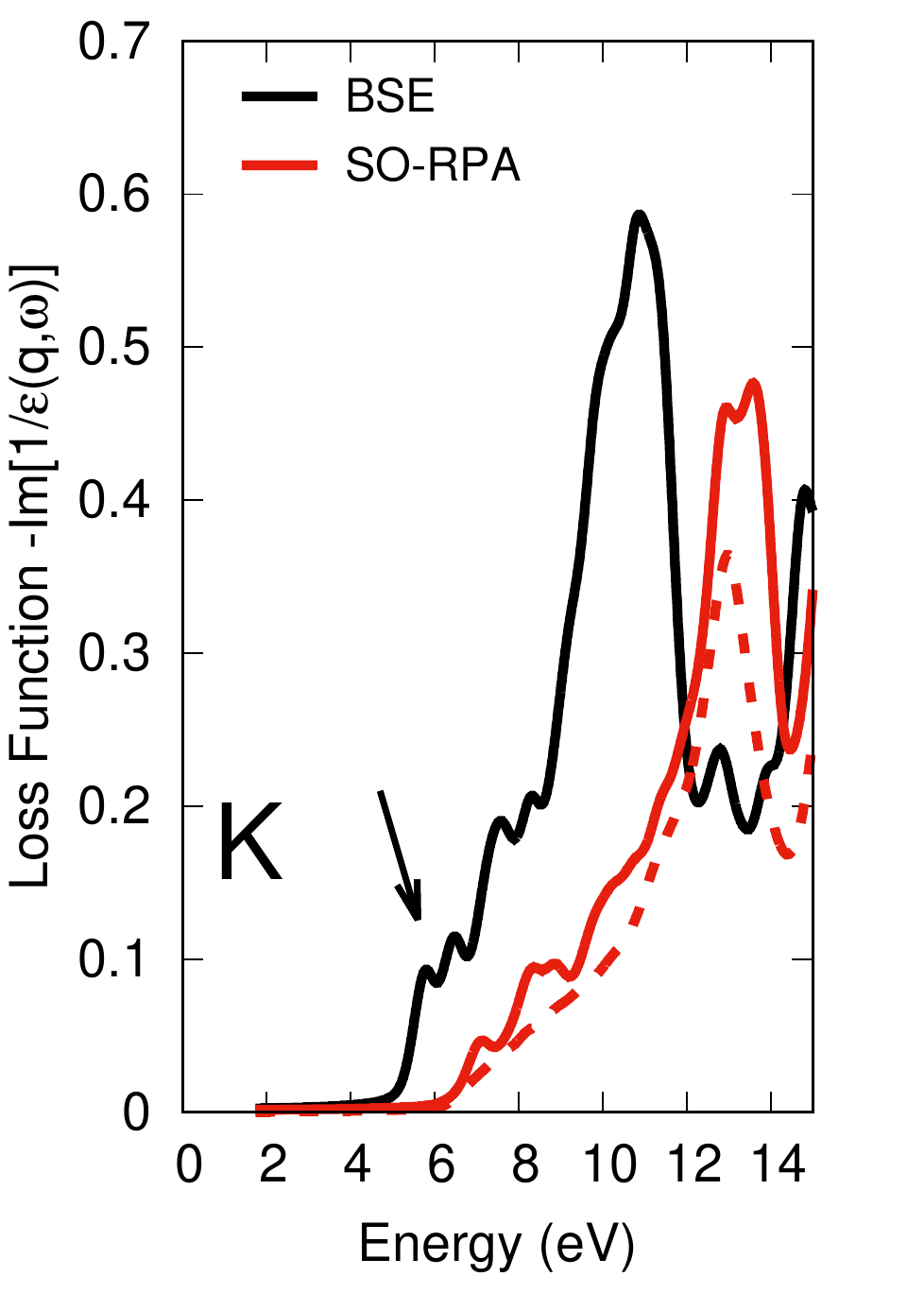}
\includegraphics[width=0.30\textwidth, trim = 20mm 55mm 160mm 20mm, clip]{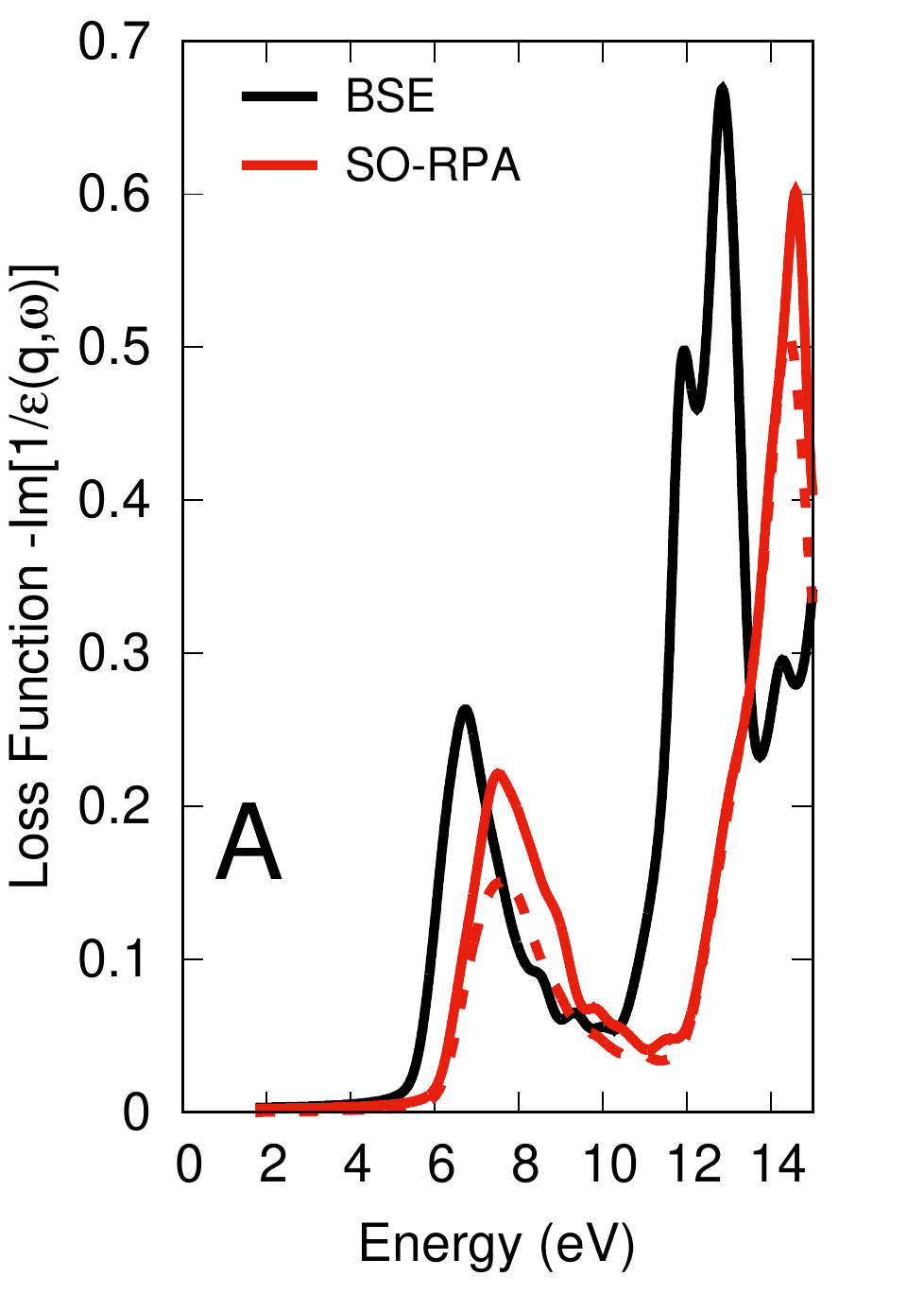}
\caption{Analysis of the excitonic structures at different momentum transfer.}
\label{fig:excitons}
\end{figure}

In $\mathbf{q}=A$ the main effect of the electron-hole interaction is a redshift of the spectral line of approximately 0.8~eV. This shift increases for higher energy, and approximately cancel the value of the scissor operator around 12~eV.
Also for $\mathbf{q}=K$ the excitonic effect is mainly a redshift of the structures, but the amount of the shift is approximately constant and similar to the scissor applied. Experimental spectra agree on the measure of a step-like onset. This feature is also predicted by BSE (arrow in Fig.\  \ref{fig:excitons}), but it is absent in RPA calculations, which clearly indicates its excitonic nature.
More peculiarities can be found in the spectrum at $\mathbf{q}=M$. Beside the usual in-plane shift already discussed, we observe here an important weight redistribution from high energy to low energy peaks. The letters in Fig.\  \ref{fig:excitons} help in tracing this redistribution.
From the comparison between the two RPA calculations, we can state that the coarseness of the k-point used in the BSE calculation can lead to a slight overestimation of the spectral intensity, however this effect is globally negligible when main characteristics are of interest. The only severe exception are peaks $D$ and $F$ at $\mathbf{q}=M$, which result excessively enhanced when compared against the calculations by Fugallo.

We continue the analysis of the structures of the loss function by discussing the plasmonic nature of the peaks.
Strictly speaking a plasmon resonance is found when $\text{Re}[\epsilon(\omega)]$ vanishes. We can adopt though a less strict definition: When a peak of the loss function is due mainly to a reduction of $\text{Re}[\epsilon(\omega)]$ (in absolute value), then the excitation has a predominant plasmonic character, when it is due to structures in $\text{Im}[\epsilon(\omega)]$, it has predominantly an inter-band transition character.
In Fig.\  \ref{fig:plasmons} we report the real and the imaginary parts of $\epsilon$ in the upper panels and the loss function in the bottom panels for $\mathbf{q}=M,K,A$.
From this analysis we can assert that the structures at 10~eV in $K$, the broad peak between 7 and 13~eV in $M$ and at 12~eV in $A$ have clearly a plasmonic nature, even though only the last one is a proper plasmon resonance. We would tend to associate a inter-band nature to the three peaks decorating the plasmon at $M$, and has a clear inter-band character the step-like onset at $K$. It is difficult instead to identify a predominant character in the structure at 7~eV at $\mathbf{q}=A$, that we would associate to some neutral excitation with mixed characteristics of a plasmonic and an inter-band excitation.

\begin{figure}[!ht]
\centering
\includegraphics[width=0.20\textwidth, trim = 10mm 15mm 160mm 15mm, clip]{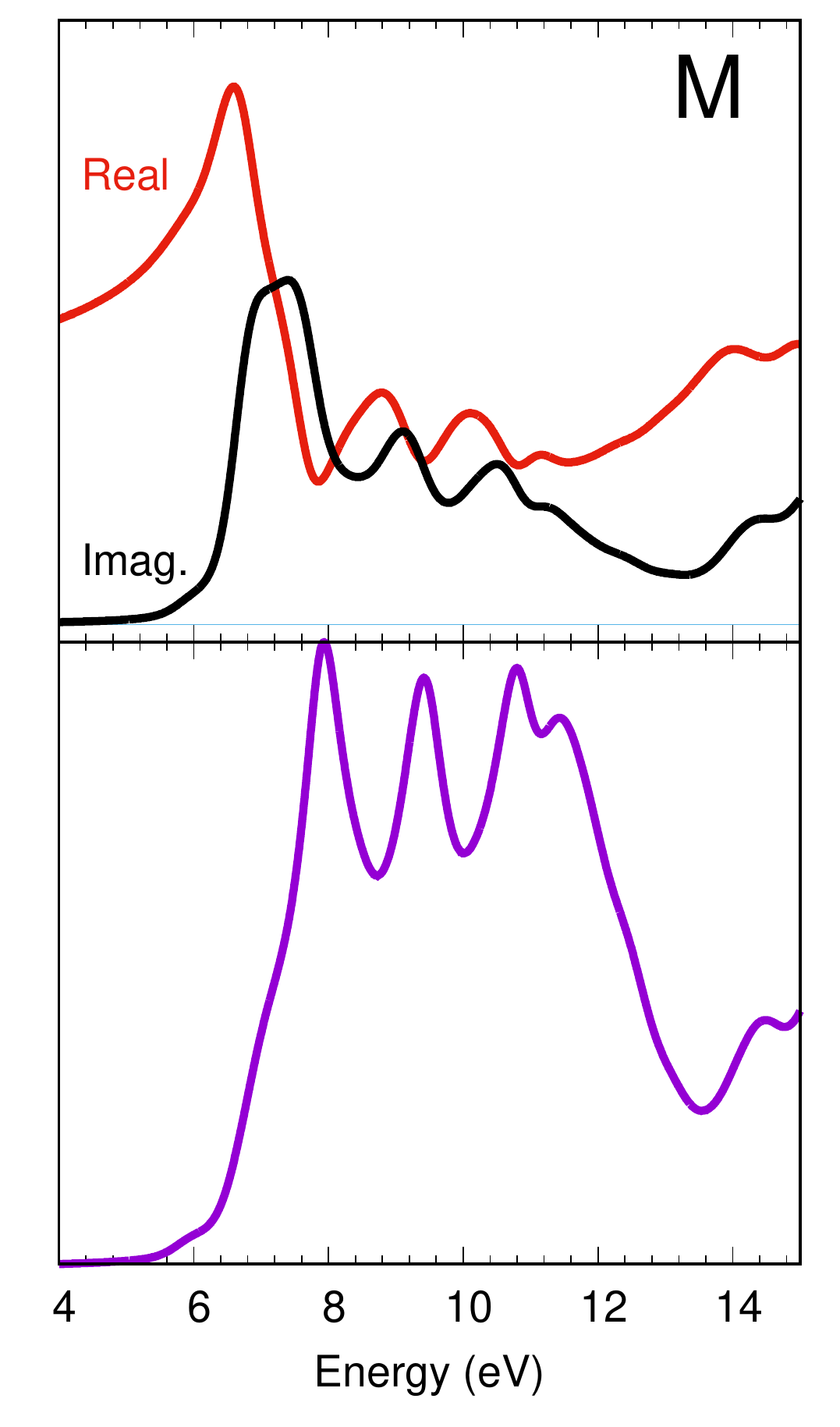}
\includegraphics[width=0.20\textwidth, trim = 10mm 15mm 160mm 15mm, clip]{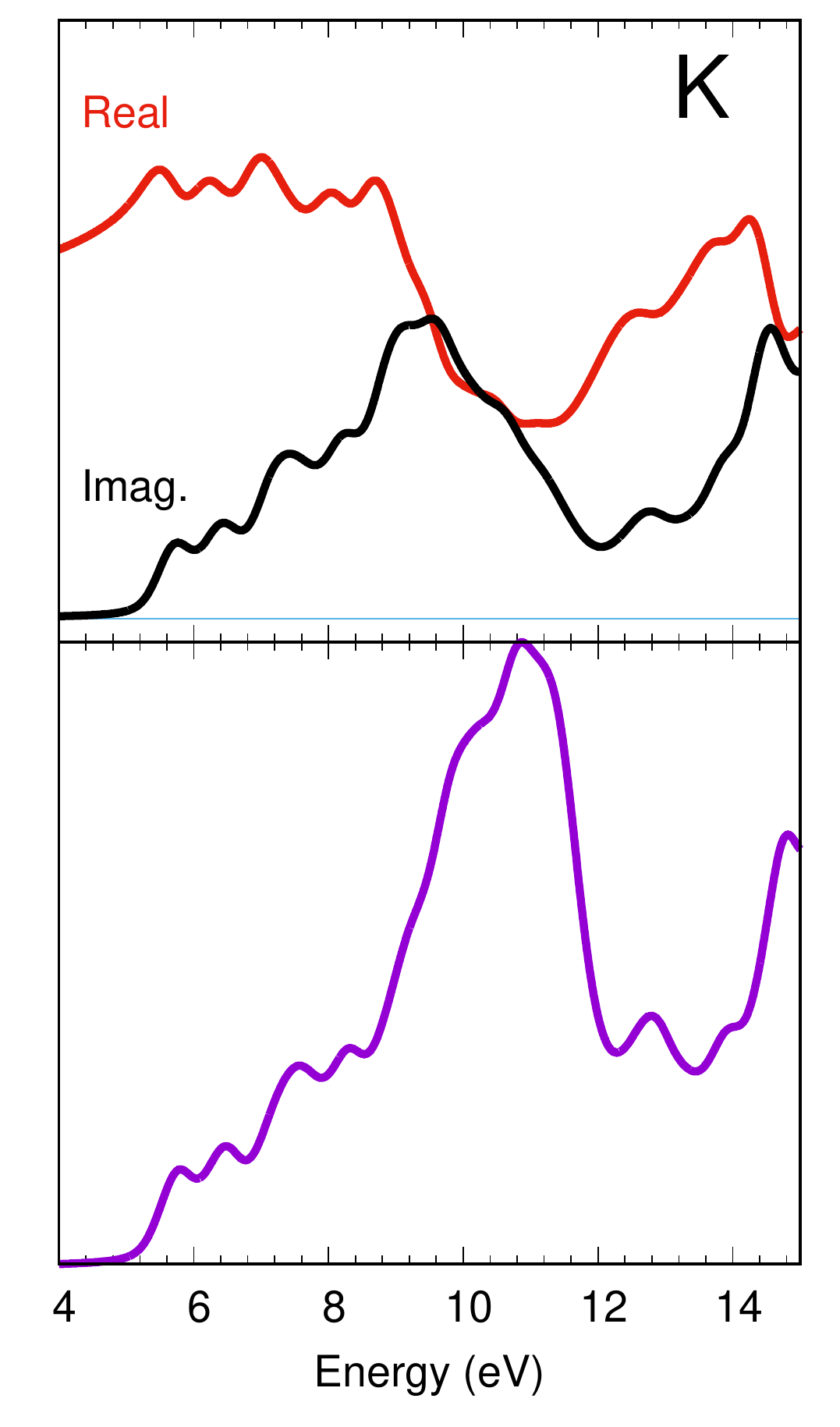}
\includegraphics[width=0.20\textwidth, trim = 10mm 15mm 160mm 15mm, clip]{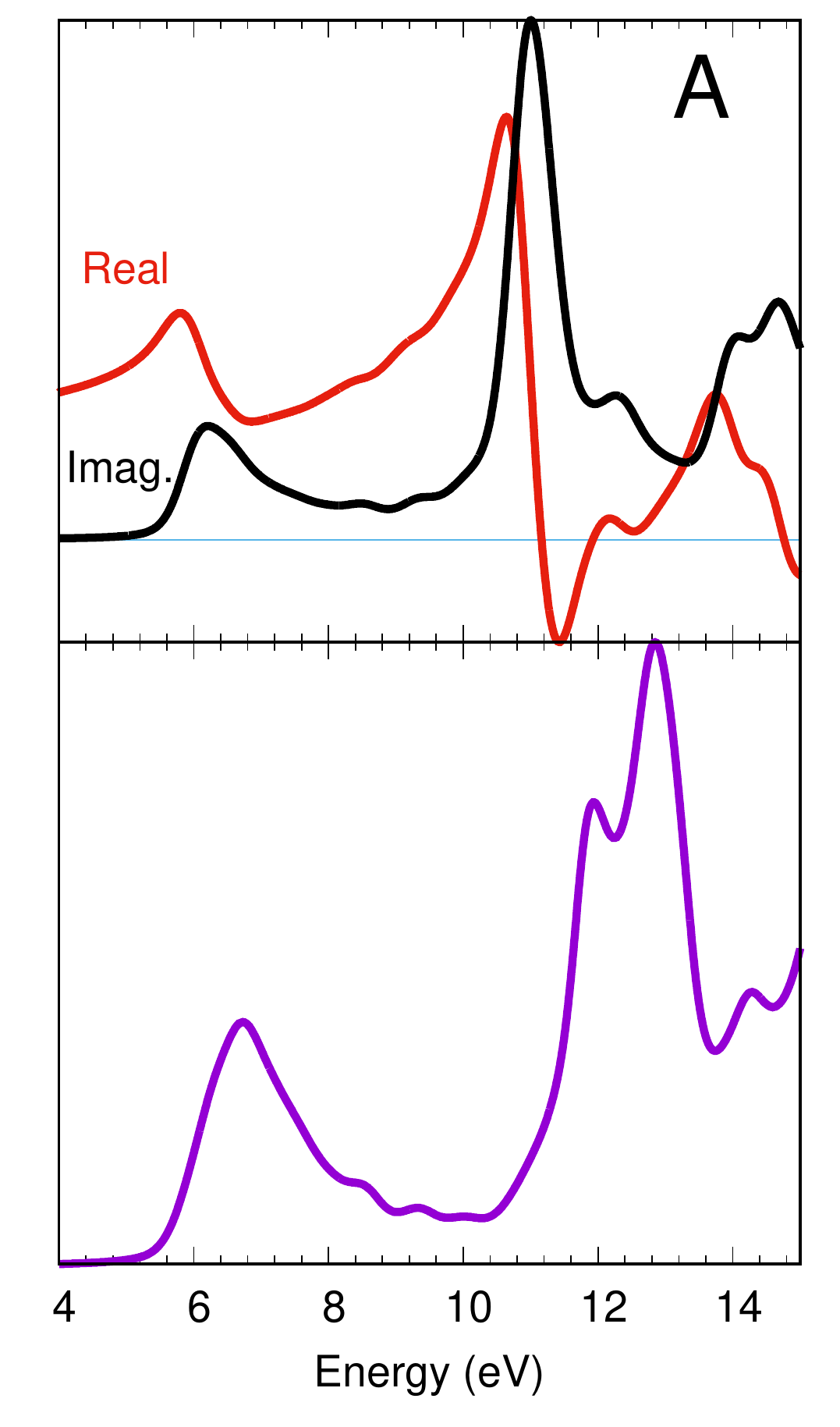}
\caption{Analysis of plasmonic excitations at different momentum transfer. Top: Real and imaginary part of the dielectric function $\varepsilon(\omega)$. Bottom: Loss function equal to -Im $\varepsilon^{-1}(\omega)$.}
\label{fig:plasmons}
\end{figure}

Finally we conclude the presentation of our theoretical results by reporting the BSE calculations at $\mathbf{q}\approx 0$ for in-plane and out-of-plane components (Fig.\  \ref{fig:plasmons2}).

\nopagebreak

\begin{figure}[!ht]
\centering
\includegraphics[width=8cm, trim = 10mm 17mm 50mm 10mm, clip]{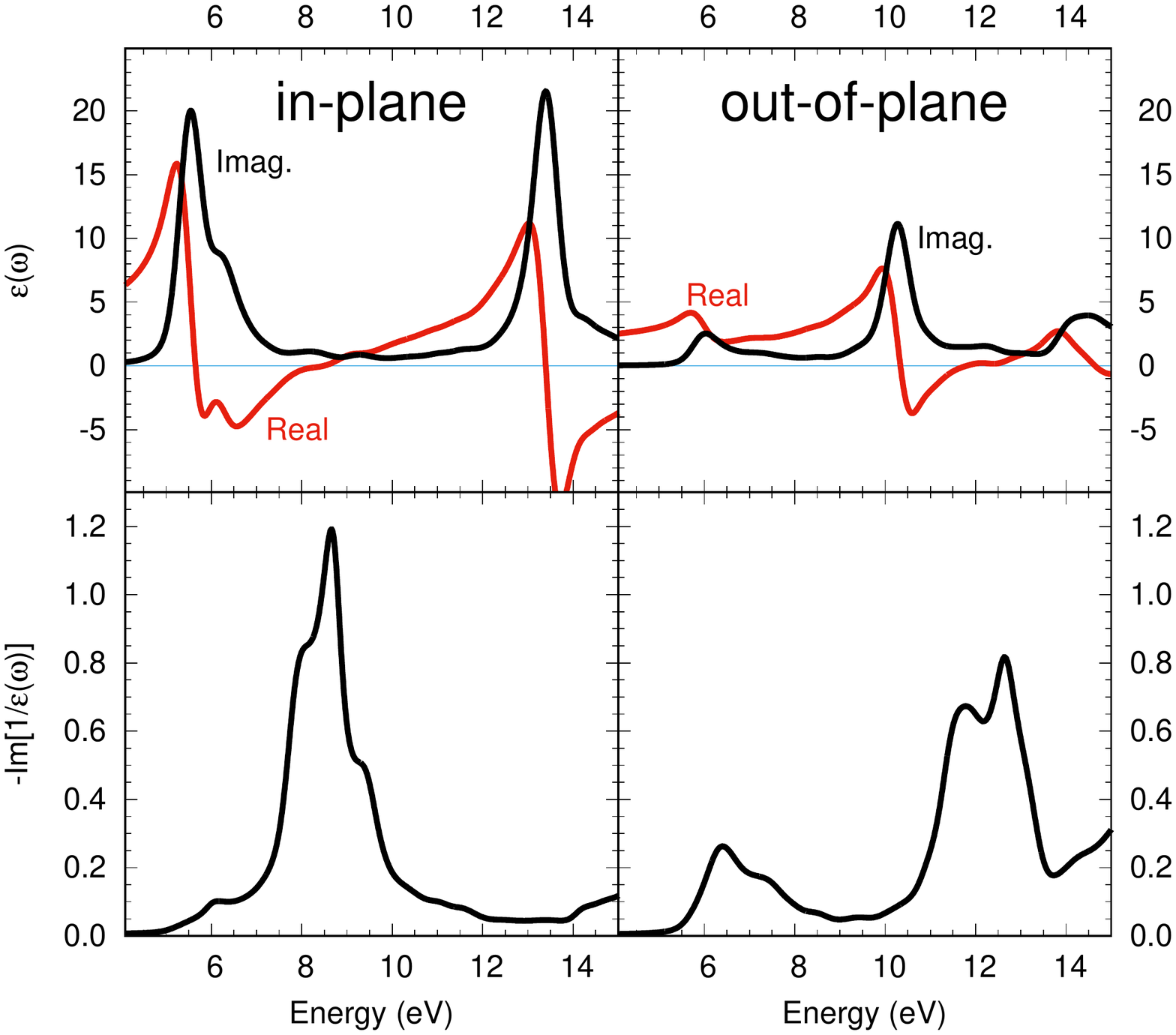}
\caption{Analysis of plasmonic excitations at $\mathbf{q}\approx 0$.}
\label{fig:plasmons2}
\end{figure}
\newpage

\end{widetext}

\end{document}